\documentclass[aps,prd,superscriptaddress,nofootinbib,twocolumn]{revtex4-2}

\usepackage{mathtools}
\usepackage{amsfonts}
\usepackage{mathrsfs}
\usepackage{comment}
\usepackage{bbm}
\usepackage{slashed}
\usepackage{amsmath}
\usepackage{bm}
\usepackage{tensor}

\usepackage{graphicx}
\usepackage{color}
\usepackage{colortbl}
\usepackage{array}
\usepackage[dvipsnames]{xcolor}

\usepackage{booktabs}
\usepackage{tabstackengine}

\usepackage{placeins}
\usepackage{makecell}
\usepackage{subcaption}
\usepackage{float}

\usepackage{xspace}
\usepackage{hyperref}
\usepackage[nameinlink]{cleveref}
\usepackage{bookmark}

\usepackage{xifthen}
\usepackage{xcolor}
\usepackage{enumitem}
\hypersetup{
	colorlinks,
	linkcolor={red!75!black},
	citecolor={blue!75!black},
	urlcolor={blue!75!black}
}
\usepackage{siunitx}

\usepackage[utf8]{inputenc}

\setkeys{Gin}{width=0.48\textwidth}
\newcommand{\tr}{{\text{tr}}}

\makeatletter
\newsavebox\myboxA
\newsavebox\myboxB
\newlength\mylenA

\newcommand*\xoverline[2][0.75]{%
	\sbox{\myboxA}{$\m@th#2$}%
	\setbox\myboxB\null
	\ht\myboxB=\ht\myboxA%
	\dp\myboxB=\dp\myboxA%
	\wd\myboxB=#1\wd\myboxA
	\sbox\myboxB{$\m@th\overline{\copy\myboxB}$}
	\setlength\mylenA{\the\wd\myboxA}
	\addtolength\mylenA{-\the\wd\myboxB}%
	\ifdim\wd\myboxB<\wd\myboxA%
	\rlap{\hskip 0.5\mylenA\usebox\myboxB}{\usebox\myboxA}%
	\else
	\hskip -0.5\mylenA\rlap{\usebox\myboxA}{\hskip 0.5\mylenA\usebox\myboxB}%
	\fi}
\makeatother

\newcommand{\tinytext}[1]{\text{\tiny{#1}}}

\newcommand{\gettitle}{Dissipation dynamics of a scalar field}

\newcommand{\getHeidelbergAffiliation}{\affiliation{Institut für Theoretische Physik, Universit\"at Heidelberg, Philosophenweg 16, 69120 Heidelberg, Germany}}
\newcommand{\getDarmstadtAffiliation}{\affiliation{Institut für Kernphysik, Technische Universit\"at Darmstadt, Schlossgartenstra{\ss}e 2, 64289 Darmstadt, Germany}}
\newcommand{\getFlorenceAffiliation}{\affiliation{
		Dipartimento di Fisica, Universit\`a di Firenze and INFN Sezione di Firenze, via G. Sansone 1,
		50019 Sesto Fiorentino, Italy
}}

\hypersetup{
	colorlinks,
	linkcolor={red!75!black},
	citecolor={blue!75!black},
	urlcolor={blue!75!black},
	pdftitle={\gettitle},
	pdfauthor={Batini, Grossi, Wink},
	pdfkeywords={critical dynamics}
	{correlations functions} {scalar theory}
	{functional renormalization group}
	{real time} {spectral function}
	bookmarksopen=true,
	bookmarksopenlevel=2,
	bookmarksnumbered=true
}

\begin{document}
	
	\title{\gettitle}
	
	\author{Laura Batini}
	\email{batini@thphys.uni-heidelberg.de}
	\getHeidelbergAffiliation
	
	\author{Eduardo Grossi}
	\email{ eduardo.grossi@unifi.it}
	\getFlorenceAffiliation
	
	\author{Nicolas Wink}
	\email{nicolas.wink@tu-darmstadt.de}
	\getDarmstadtAffiliation

	\begin{abstract}
		We investigate the dissipation rate of a scalar field in the vicinity of the phase transition and the ordered phase, specifically within the universality class of model A. This dissipation rate holds significant physical relevance, particularly in the context of interpreting effective potentials as inputs for dynamical transport simulations, such as hydrodynamics.
		To comprehensively understand the use of effective potentials and other calculation inputs, such as the functional renormalization group, we conduct a detailed analysis of field dependencies.
		We solve the functional renormalization group equations on the Schwinger-Keldysh contour to determine the effective potential and dissipation rate for both finite and infinite volumes.
		
		Furthermore, we conduct a finite-size scaling analysis to calculate the dynamic critical exponent $z$. Our extracted value closely matches existing values from the literature.
	\end{abstract}
	
	\maketitle

	\section{Introduction}
	
	Second-order phase transitions are fundamental phenomena in physics, occurring in a wide range of microscopically different classical and quantum systems.
	While phase transitions in equilibrium are comparatively well studied, transitions in nonequilibrium systems present a significant challenge and remain much less explored. 
	Nonequilibrium phase transitions are particularly interesting as they exhibit unique dynamical properties.
	
	In the context of quantum chromodynamics (QCD) and heavy-ion collisions, this is closely related to the search for a potential critical end point.
	Utilizing the classification of dynamical universality classes of Hohenberg and Halperin \cite{Hohenberg:1977ym}, it is believed that the dynamics of the critical point is captured by model H. Although the study of model H remains challenging, it is interesting to study simpler models such as model C (see, e.g.,~\cite{Berges:2009jz, Berges:2012ty, Schweitzer:2020noq, Roth:2023wbp}), which includes coupling to the conserved energy density.
	
	A well-known example of nonequilibrium universality is the critical relaxational dynamics of statistical systems in contact with a thermal reservoir, known as model A. It corresponds to the Glauber dynamics of the Ising model, which is perhaps the simplest case of critical dynamics, characterized by the absence of conservation laws. This model is even simpler than model C but is sufficient for our study, described in detail below. For studies of this model, see, e.g.,~\cite{Hohenberg:1977ym, prudnikov1997critical, nightingale2000monte, Canet:2006xu, Canet:2011wf, mesterhazy2015quantum, duclut2017frequency, Zhong:2018yjx, hasenbusch2020dynamic, Schweitzer:2020noq, Huelsmann:2020xcy, adzhemyan2022dynamic, Schaefer:2022bfm, Roth:2021nrd, Florio:2021jlx, Florio:2023kmy, Roth:2023wbp, Chao:2023kvz}.
	
	Near second-order phase transitions, critical slowing down occurs, indicating that the dynamics becomes significantly slower on mesoscopic time scales near the phase transition. Furthermore, in many cases, the computation of evolution equations for the density matrix becomes impractical due to the involvement of many degrees of freedom. As a result, it is often more convenient to describe the system's properties using a mesoscopic variable, which considers observables that dictate the system's behavior on length and time scales larger than the microscopic ones.
	For example, this mesoscopic variable could be a classical field, such as the coarse-grained local magnetization field in a magnetic system. The effective Hamiltonian, formulated in terms of this field, provides a comprehensive description of the system's properties.
	
	To investigate the near-equilibrium evolution of coarse-grained systems, transport or hydrodynamic simulations are often utilized, see, e.g., \cite{Herzog:2008he, Peralta-Ramos:2011ilx, Glorioso:2016gsa}. 
	In this context, only a limited set of variables, such as temperature and velocity, is considered. The equation of motion in fluid dynamics often simplifies the density conservation, reflecting the system's microscopic symmetries.
	Near a second-order phase transition, it is natural to include the order parameter as a state variable with temperature and velocity; thus, the usual fluid-dynamic equation must be extended to include the order parameter. The equation of motion of model A is the simplest effective equation of motion that can appear since the order parameter is not conserved and is decoupled from any conserved charge.
	However, hydrodynamic methods typically rely on classical Hamiltonians, which can limit their ability to capture the system's behavior accurately. Instead, it is preferred to use an effective coarse-grained Hamiltonian, which describes the system's dynamics more accurately on a larger scale.
	In this work, we discuss the usage of coarse-grained input, such as the effective potential, in transport and hydrodynamic simulations.
	On a technical level, this concerns the compatibility of the convexity of the effective potential with such evolution equations and the dissipation dynamics in its vicinity.
		
	We employ the functional renormalization group (fRG) (see, e.g.,~\cite{Dupuis:2020fhh} and references therein) to study model A. Previously, model A has been the subject of investigation using the fRG technique in~\cite{Canet:2006xu, Canet:2011wf, mesterhazy2015quantum, duclut2017frequency, Huelsmann:2020xcy, Roth:2021nrd, Roth:2023wbp}.
	This approach allows us to investigate coarse-grained quantities and provides a unified framework for describing both the static and dynamic properties of phase transitions, as well as the nonuniversal behavior of the system.
	Methods such as the fRG could provide valuable first-principle input to close the resulting system of equations if the aforementioned difficulties are overcome. A comprehensive description of the process involved in achieving this outcome stands as one of the main results of this work.
	
	The paper is organized as follows.
	In Secs.~\ref{sec:Hydro} and \ref{sec:CriticalPhenomena}, we introduce model A, its connection to hydrodynamics, and how we gain access to the dissipation rate and the effective potential within the fRG.
	The results are collected in Secs.~\ref{sec:crit_dynamics} and~\ref{sec:non_crit_results}, focusing on critical and noncritical aspects, respectively.
	Finally, we conclude our work with a summary and discussion in Sec.~\ref{sec:Conclusion}.
	
	To aid the reader in understanding the details of our work, we provide several appendices that cover various aspects related to model A formulated on the closed time path and fRG flows, including flows in finite spatial volumes. These Appendices (Appendices~\ref{app:dyn_field_theory} and \ref{app:flows}) are intended to provide additional context and insight into the methods employed in our study. Additionally, we include a detailed account of our numerical implementation in Appendix~\ref{sec:num_alg}.
	
	The code used to produce the results in this work is available on \href{https://github.com/laurabatini/flow-equations-code}{GitHub}~\cite{Githubcode}.
	\section{Hydrodynamic description}
	\label{sec:Hydro}

	A standard procedure for describing dynamic critical phenomena is identifying the relevant hydrodynamic degrees of freedom. This approach is favored due to the computational infeasibility of determining the density matrix's evolution equations, which involve many degrees of freedom.
	In a hydrodynamic sense, the system we are considering can be divided into microscopic (fast) internal degrees of freedom that behave like a bath, which we assume to be thermal equilibrium coupled to the more macroscopic (slow) dynamics of the degrees of freedom. 
	The fluid dynamics approach allows for slow variations of temperature, pressure, and velocity in space and time.
	Furthermore, in the vicinity of a second-order phase transition, the order parameter becomes nonvanishing and, therefore, needs to be included in the hydrodynamic theory as an additional state variable along with other fields such as temperature and velocity. 
	Consequently, the conventional hydrodynamic equation must be extended to include the dynamics of the order parameter. 
	
	The derivation of model A in terms of fluid-dynamics considerations is mostly unknown. Therefore, we introduced the model using entropy considerations to highlight the similarity to the standard fluid-dynamics theory~\cite{landau2013statistical}. A more standard formulation can be found in~\cite{tauber2014} and Appendix~\ref{app:dyn_field_theory}. 
	
	Here, we first present the hydrodynamic theory in the presence of an order parameter in the ideal case according to~\cite{Grossi:2021gqi, Grossi:2020ezz}. 
	We then go on to discuss a more realistic case where the conservation laws are only statistically preserved and locally perturbed. We describe how, in such a scenario, viscous effects enter the evolution of the spatially averaged energy-momentum tensor and the evolution equation of the fluid field. 
	This modified description represents the simplest effective equation of motion since the order parameter is not conserved and does not interact with other conserved charges.
	
	\subsection{Ideal fluid} \label{subsec:Idealhydro}

	We adopt a practical procedure outlined in \cite{Jensen:2012jh} to derive the ideal hydrodynamic equations. 
	These equations are obtained from the hydrodynamic action given by
	\begin{equation}
		S\left[g_{\mu \nu}\right]=\int d^4 x \sqrt{-g} \, p_{\phi}(T,\left(\partial_{\perp} \phi\right)^2, \phi ) \, .
	\end{equation}
 Here, the pressure $ p_\phi$ is defined as
\begin{equation}
	p_\phi(T, (\partial_\perp \phi)^2 , \phi) \equiv p_0(T)  -U(\phi) -  \frac12 \Delta^{\mu\nu} \partial_{\mu}\phi \cdot  \partial_{\nu}\phi \, ,
	\end{equation}
  and depends on the temperature $T \equiv (-\beta^{\mu}g_{\mu\nu} \beta^{\mu})^{-1/2}$, the fluid velocity $u^\mu \equiv T \beta^\mu$, and the order parameter field $\phi$. 
	Here, $ \partial^\mu_{\perp} = \Delta^{\mu \nu} \partial_{\nu}$, $\Delta^{\mu \nu} = g^{\mu \nu}+ u^\mu u^\nu$ represents the projection operator orthogonal to the fluid velocity, and $p_0$ is the ideal pressure for vanishing order parameter. 
	We have an explicit dependence on the metric $g_{\mu \nu}$ to derive the stress tensor, and ultimately this source will be set to zero.
	The conserved energy-momentum tensor  $T^{\mu\nu}_\tinytext{ideal}$, derived by varying the action
	 with respect to the metric $g_{\mu \nu}$, can be parametrized in the Landau frame as 
	\begin{equation}
		\begin{aligned}
			T^{\mu\nu}_\tinytext{ideal} = &   \left. \frac{2}{\sqrt{-g}}\frac{\partial S}{\partial g_{\mu \nu}} \right|_{g=0} =  
			\, (\varepsilon_\phi +p_\phi)\,u^{\mu}u^{\nu} +p_\phi g^{\mu\nu} \\ & + 
			\partial^{\mu}\phi\cdot \partial^{\nu}\phi  -u^\mu u^{\nu} (u^\sigma \partial_{\sigma} \phi)\cdot (u^\rho \partial_{\rho}\phi) \, .
		\end{aligned}
		\label{eq:energy_mom_tensor_ideal}
	\end{equation}
	The last two terms in \labelcref{eq:energy_mom_tensor_ideal} involving the field are the energy and momentum contributions coming from the field, respectively.
	The energy density $\varepsilon_{\phi}$ is defined through the Legendre transform of the pressure $p_{\phi}$ as
	\begin{equation}\label{redef-energy}
		\varepsilon_\phi  \equiv 	\left(-1+T\frac{\partial  }{\partial T}  	\right) p_\phi \, .
	\end{equation}
	The corresponding conservation law $\partial_{\mu} T^{\mu\nu}_\tinytext{ideal}=0$ leads to the ideal effective equation for the order parameter \cite{Grossi:2021gqi,Grossi:2020ezz}
	
	\begin{equation}
		\label{Josephsonconstrain}
		u^{\mu} \partial_{\mu} \phi = 0 \, ,
	\end{equation}
	provided that the field is in the minimum of the static free energy.
	Therefore, the dynamics neglecting dissipation is trivial:
	the field remains at its minimum and simply moves with the fluid velocity. To introduce more complex dynamics, it becomes essential to incorporate dissipative effects. These effects enable the field to deviate from the free energy minimum, leading to nontrivial behavior.
	
	\subsection{Dissipative fluid} \label{subsec:Vischydro}
	So far, we have only considered the ideal equations of motion. In the presence of dissipation, the energy-momentum tensor expressed in \labelcref{eq:energy_mom_tensor_ideal} is modified as follows:
	\begin{equation}
		T^{\mu\nu} = T^{\mu\nu}_{\tinytext{ideal}} + \Pi^{\mu \nu}.
		\label{eq:dissipative_correction}
	\end{equation}
	Here, we continue working in the Landau frame, ensuring the stress tensor $\Pi^{\mu \nu}$ satisfies $\Pi^{\mu \nu} u_\mu = 0$. The stress tensor can be decomposed as:
	\begin{equation}
		\Pi^{\mu \nu} = \pi_{\tinytext{bulk} } \Delta^{\mu\nu} + \pi^{\mu \nu},
	\end{equation}
	where $\pi^{\mu \nu}$ represents the shear stress tensor, satisfying $\pi_{\mu}^{\mu} = u_\mu \pi^{\mu \nu} =  0$.
	
	By utilizing the conservation of the energy-momentum tensor, the Gibbs-Duhem relation \labelcref{redef-energy}, and the pressure differential, given by
	\begin{equation}
		d p_{\phi}=s_{\phi} d T-\frac{1}{2} d\left(\partial_{\perp}^\mu \phi\right)^2-\frac{\partial U}{\partial \phi} d \phi \, ,
	\end{equation}
	we can derive the entropy production as 
	\begin{align}
		\partial_\mu(s_\phi u^{\mu} )= 
		\frac{\Xi}{T} \Theta
		- \partial_{\mu}\left(\frac{u_{\nu}}{T}\right)\Pi^{\mu\nu}
		\, ,
	\end{align}
	where we have defined the scalar quantities
	\begin{equation}
		\Theta  \equiv \partial^2_{\perp}\phi -\frac{\partial U}{\partial \phi } \, , \quad
		\text{and} \quad
		\Xi \equiv u^{\mu}\partial_\mu\phi
		\, .
	\end{equation}
	
	Besides the dissipative corrections to the energy-momentum tensor, the evolution equation of the order parameter $\phi$ expressed in \labelcref{Josephsonconstrain} is modified by dissipative effects. 
	By including the first-order corrections in the gradient expansion of hydrodynamics, the shear stress tensor can be written as
	\begin{equation}
		\pi^{\mu\nu} = -\eta \sigma^{\mu\nu} ,
	\end{equation}
	where $\eta$ is the shear viscosity. The requirement of positive entropy production in the tensor sector can be ensured by imposing $\eta \ge0$.
	For the scalar sector, the terms can be expressed as
	\begin{align}
		\pi_{\tinytext{bulk}}&=-\zeta \, \partial_\mu u^{\mu} - \zeta^{(1)} \, \phi \Theta \, ,\\
		\Xi&= \zeta^{(1)}\, \phi \partial_\mu u^{\mu}+ \Gamma \, \Theta \, ,  
	\end{align}
	where $\zeta$ denotes the well-known bulk viscosity, and  $\Gamma$ is the transport coefficient regulating the dissipative effects of the scalar field dynamics, respectively. 
	The coefficient $\zeta^{(1)} $ is an independent transport coefficient that couples the expansion rate $\partial_\mu u^{\mu}$ to the relaxation equation for the field and vice versa.
	To ensure the positivity of the associated quadratic form, the following conditions must be satisfied:
	\begin{equation}
		\zeta \ge0 \, ,\quad \Gamma\ge0 \, , \quad \text{ and } \quad \zeta \,\Gamma - (\zeta^{(1)})^2\, \phi^2 \ge 0 \, .  
	\end{equation}
	By specifying the dissipative fluxes, it becomes possible to formulate the scalar field equation~\cite{landau2013statistical}. The scalar field obeys a dissipation-type equation  
	\begin{equation}\label{josephson-diss}
		u^{\mu}\partial_{\mu}\phi = \Gamma \left[
		\partial^2_{\perp}\phi -\frac{\partial U}{\partial \phi} 
		\right] +\zeta^{(1)} \phi \, \partial_\mu u^{\mu}. 
	\end{equation}
	In the limit of zero velocity, the equation of motion for the field reduces to
	\begin{equation}\label{josephson-diss2}
		\partial_{t}\phi = \Gamma \left[
		\nabla^2\phi -\frac{\partial U}{\partial \phi} 
		\right]
		\, .
	\end{equation}
	This equation has to be intended as the effective evolution equation of the expectation value of the scalar field. 
	As usual, in the high-temperature limit, the stochastic noise $\xi$ has to be introduced on the right-hand side 
	of \labelcref{josephson-diss2}
	\begin{equation}\label{josephson-diss3}
		\partial_{t}\phi = \Gamma \left[
		\nabla^2\phi -\frac{\partial U}{\partial \phi} 
		\right]  + \xi \, ,
	\end{equation}
	with $\langle \xi \rangle =0 $ and 
	$\langle \xi(x)\xi(x^{\prime}) \rangle = 2T \Gamma \delta(x-x^{\prime })$  such that the fluctuation-dissipation theorem holds~\cite{landau2013statistical, Petrosyan:2021lqi}.

	From the previous analysis, it is clear that $\Gamma$ can be treated similarly to other first-order transport coefficients.
	
	The main goal of this paper is to calculate a specific transport coefficient, the dissipation rate represented as $X = \Gamma^{-1}$, and study its dependence on both temperature and the field expectation value. It is important to note that in the vicinity of the phase transition, where the order parameter $\phi$ approaches zero, the last term on the right-hand side of the equation becomes negligible and can be safely ignored. However, when the system is situated far from the phase transition, this term becomes substantial and should be accounted for in the analysis.
	
	To avoid confusion, it is important to note that the symbol $\Gamma$ is also commonly used to denote the quantum effective action. Therefore, we will use the symbol $X$ exclusively to refer to the dissipation rate while reserving the symbol $\Gamma$ to denote the average effective action, cf.~\labelcref{eq:effGamma}.
	
	\section{Theoretical setup}  \label{sec:CriticalPhenomena}
	In the previous section, we presented the effective dynamics of the order parameter at the mesoscopic scale within the hydrodynamics framework derived through considerations of entropy production.
	In this section, we turn towards the description of model A within the fRG formalism.
	
	\subsection{Model A}
	In model A, the order parameter is represented by a coarse-grained single-component real scalar field $\phi(t,\boldsymbol{x})$, which relaxes towards its equilibrium value over time. 
	The nonlinear Langevin equation governs the effective dynamics of the order parameter (see, e.g.,~\cite{Canet:2006xu})
	\begin{equation}
		\label{eq:langevin_evolution}
		\partial_t \phi(t,\boldsymbol{x})=-\frac{1}{X} \frac{\delta \mathcal{H}[\phi]}{\delta \phi(t,\boldsymbol{x})}+\xi(t,\boldsymbol{x})
		\, ,
	\end{equation}
	where $\mathcal{H}[\phi]$ represents the Hamiltonian, and $t$ and $\boldsymbol{x}$ denote the time and space coordinates, respectively.
	The coefficient $1/X$ represents a constant and uniform dissipation rate, and $\xi(\mathbf{x}, t)$ is a stochastic white noise with the following properties:
	\begin{equation}
		\begin{aligned}
			\langle\xi(t,\boldsymbol{x})\rangle &=0 \, , \\
			\left\langle\xi(t,\boldsymbol{x}) \xi\left(t^{\prime},\boldsymbol{x}^{\prime}\right)\right\rangle &=2 \mathcal{N} \delta^{(d+1)}(x-x^\prime) 
			\, ,
		\end{aligned}
	\end{equation}
	where $\mathcal{N}$ is a constant that quantifies the thermal fluctuations induced by the heat bath at temperature $T$, and $d$ is the number of spatial components. In the long-time limit, the system is stationary, and consequently, the noise is assumed to satisfy the detailed balance condition $\mathcal{N} = T X$~\cite{tauber2014}.
	We choose the deterministic drift in the model to be the $\mathcal{Z}_2$-invariant Landau-Ginzburg effective Hamiltonian 
	\begin{equation}
		\mathcal{H}[\phi]=\int d^d x\Big\{\frac{1}{2}[\nabla \phi(x)]^2+U(\phi)\Big\}
		\, , 
		\label{eq:Isinghamiltonian}
	\end{equation}
	with the potential term given by
	\begin{equation} 
		U(\phi)=\frac{m^2}{2} \phi^2+\frac{\lambda}{4} \phi^4,
		\label{eq:phi4potential}
	\end{equation}
	where  $m$ represents the mass parameter. 
	When $m^2 < 0$, the system is in the ordered or broken phase, while when $m^2>0$, it is in the symmetric phase.
	The parameter $\lambda$ represents an interaction term that arises from the nearest-neighbor interactions in the underlying lattice Ising model that gives rise to the field theory. To ensure stability, it is necessary to choose $\lambda > 0$. 
	
	To derive the associated field theory from the Langevin equation, the Martin–Siggia–Rose–Janssen–De~Dominicis procedure~\cite{Martin:1973zz, dominicis1976techniques, janssen1976lagrangean, DeDominicis:1977fw} has been introduced and widely used in the literature. For a pedagogical introduction to this procedure, see, e.g.,~\cite{Canet:2011wf, tauber2014, Hertz:2016vpy}.
	
	See Appendix~\ref{app:dyn_field_theory} for a detailed description of the Schwinger-Keldysh contour and the derivation of the flow equations using the Schwinger-Keldysh approach in a dynamical field theory. 
	Conveniently, the Keldysh formalism has the advantage that it can be successfully applied to flowing coupling constants in the context of the renormalization group without significant complications. 
	By doing so, one obtains an infinite system of coupled integral-differential equations that must be truncated. The action can be expanded using the derivative expansion, i.e., at a certain order of space-time derivatives of the fields. 
	This work focuses on the derivative expansion, as it captures all relevant dynamics of interest. A detailed account of the resulting ansatz for the effective action is given in the following subsection.
	\subsection{Ansatz for the effective action}
	\label{app:modelA}
	To accurately capture the dissipation dynamics, we adopt an expansion that includes all relevant effects for understanding the system. 
	In our approach, we consider a leading-order expansion of the effective action $\Gamma_k $ in gradients of the field $\phi_r$, introduced below, and retaining the complete field dependence.
	Specifically, we truncate at order one in the time derivative and second order in space derivatives and obtain as the ansatz for the effective action
	\begin{equation}
		\label{eq:effGamma}
			\Gamma_k[\Phi]=  \int_{t,\boldsymbol{x}}\phi_a \Big(
			X_k(\phi_r)(\partial_t \phi_r- i \phi_a)
			- \nabla^{2} \phi_{r}+ U_k^{(1)}(\phi_r)\Big),
	\end{equation}
	where now $\Phi$ is the duplet of fields $\Phi = (\phi_r, \phi_a)$, cf.~\labelcref{eq:Keldysh_basis}, and the indices at the integral sign simply refer to an integration over space-time.
	The parametrization of the effective action is chosen such that the fluctuation-dissipation theorem holds at all RG scales $k$; see Appendix~\ref{app:dyn_field_theory} for more details.
	
	At this level of approximation, there are only two scale-dependent functions, given by the first derivative of the effective potential $U^{(1)}_k(\phi) = \partial U_k(\phi) /\partial \phi$ and the dissipation rate $X_k(\phi)$. Their flows are obtained by suitable projections of the effective action
	\begin{equation}
		\begin{aligned}
			&U^{(1)}_k (\phi) =  \frac{1}{\text{Vol}_{d+1}}\Bigg(\frac{\delta \Gamma_k[\Phi]}{\delta \phi_a(q)}\Bigg|_{\Phi = ( \phi, 0)}\Bigg)\Bigg|_{\nu=0, \boldsymbol{q}=0}, \\
			&	X_k (\phi) = \frac{1}{\text{Vol}_{d+1}}\partial_{i \nu}\Bigg( \frac{\delta^{2} \Gamma_{k}[\Phi]}{\delta \phi_r(q) \delta \phi_a(-q)}\Bigg|_{\Phi = ( \phi,0)}\Bigg)\Bigg|_{\nu=0, \boldsymbol{q}=0}
			\, ,
		\end{aligned}
	\end{equation}
	where $\Phi(t,\boldsymbol{x}) = (\phi,0)$ is a uniform and stationary field configuration. 
	
	Contrarily to perturbation theory, the couplings are arbitrary functions of the field. This requirement is essential if we want to study the theory in the phase with broken symmetry and obtain the full effective potential, as it can be nonanalytic.
	
	
		\begin{figure*}[t!]
		\centering
		\begin{subfigure}[t]{0.48\textwidth}
			\centering
			\includegraphics[width=\linewidth]{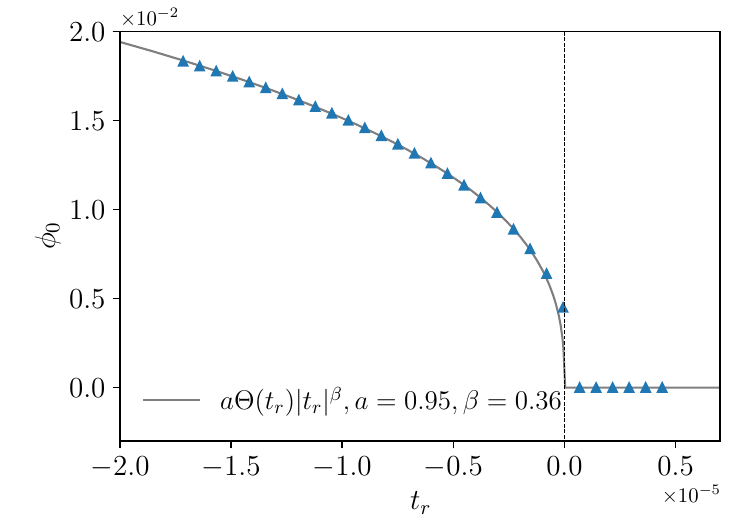}
			\label{fig:mass_crit}
		\end{subfigure}	~
		\begin{subfigure}[t]{0.48\textwidth}
			\centering
			\includegraphics[width=\linewidth]{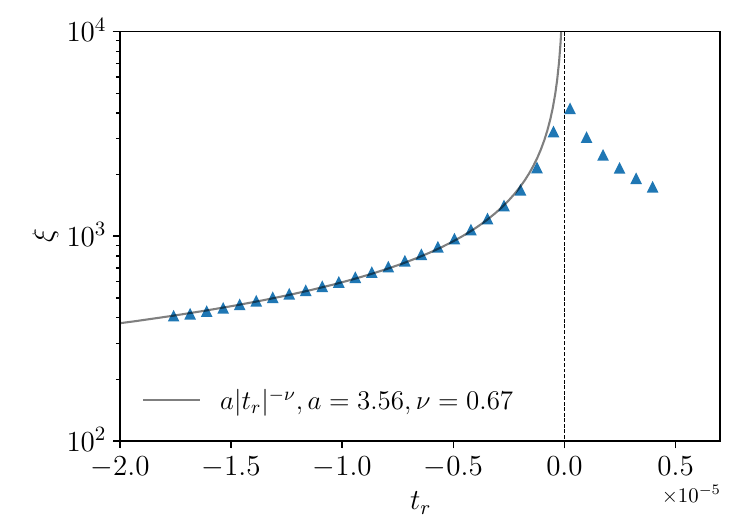}
			\label{fig:Uprimeprime}
		\end{subfigure}%
		\caption{ Field expectation value $\phi_0$ and correlation length $\xi$ across the phase transition with critical mass $m^2_c = -0.927814$, in the mass range $m^2\in [-0.927830,-0.927810]$.
			The numerical simulation result is shown with blue triangles, and the corresponding fits are shown by a solid line.
			\textbf{Left:}	Field expectation value $\phi_0$ as a function of the reduced temperature $t_r$ \labelcref{eq:redtemp} fitted with the leading-order scaling ansatz $	\phi_0(t_r)=a \Theta(t_r) |t_r|^\beta$ and $a=0.95$, $\beta = 0.36$.
			\textbf{Right:} Correlation length $\xi $ as a function of the reduced temperature $t_r$ fitted on the left with the leading order scaling ansatz $\xi = a |t_r|^{-\nu},$ with $a=3.56$, $\nu= 0.67$. 	}
		\label{fig:Critical_fits}
	\end{figure*}
	
	\subsection{Renormalization-group equations } \label{subsec:RG-equations}
	The set of flow equations for the running couplings is derived in detail in Appendix~\ref{app:flows}.
	Their flow follows from the flow of the one- and two-point functions, derived from the flow equation \labelcref{eq:wetterich} and moments of the quantum effective action $\Gamma_k$.
	In order to ease the notation we will, from now on, drop the subscripts indicating the scale $k$.
	
	Furthermore, the couplings exhibit field dependence, as do their derivatives, but we will drop this dependence from now on.
	The well-known flow equation for the derivative of the effective potential is given by [from \labelcref{eq:flow_effpot}]
	\begin{align}
		\partial_k U^{(1)}= \frac{\Omega_d}{(2 \pi)^d d}  \frac{\partial}{\partial \phi} \left(   \frac{k^{d+1}}{U^{(2)}+ k^2  }
		\right)
		\, ,
		\label{eqforV}
	\end{align}
	where $\Omega_d = 2 \pi^{d/2} \Gamma(\frac{d}{2})^{-1}$ is a factor resulting from the angular integrations and $\Gamma$ is the gamma function.
	
	The equation for the dissipation rate $X$ \labelcref{eq:flow_} reads
	\begin{equation}
		\partial_k X = -\frac{\Omega_d k^{d+1}}{2 (2\pi)^d}\Big[3 (\partial_\phi G)^2 X + 4\partial_\phi (G^2)X^{(1)}+ 2 G^2X^{(2)}\Big]
		\, ,
		\label{eqforX}
	\end{equation} 
	where, for the sake of convenience, we have set 
	\begin{equation}
		\label{definitionG}
		G = \frac{1}{k^2 + U^{(2)}}
		\, .
	\end{equation} 
	Note that the dissipation rate $X$ does not affect the equation for the effective potential $U$, which is the standard equilibrium flow equation of the Ising model.
	This is not surprising because model A satisfies for any scale $k$ the fluctuation-dissipation relation, which is the hallmark of thermal equilibrium.
	Consequently, the critical exponents $\nu$ and $\eta$ for model A are the same as in the static Ising model.
	However, the converse is not valid; the equation for the dissipation rate depends on $U$ and its derivatives. As such, it is necessary to solve the flow equations as a system of coupled partial equations. Further details regarding the numerical implementation of the flow equations can be found in Appendix~\ref{sec:num_alg}.

	To initialize the system, we specify the couplings at the UV initial RG scale $k= \Lambda = 10$ as
	\begin{equation}
		U^{(1)}_{k= \Lambda}(\phi)= m^2\phi +  \lambda \phi^3, \quad  X_{k = \Lambda}(\phi)= 1 \,. 
	\end{equation}
	This allows us to use (in the UV) the mass variable replacing the temperature. 
	Indeed, if the mass value exceeds a critical threshold, the system will end up in the broken phase in the IR. In other words, this situation is analogous to working at temperatures below a critical temperature ($T < T_c$).
	Conversely, the system will end up in the symmetric phase in the IR, equivalent to working at temperatures $T>T_c$. 
	Additionally, we use our freedom to fix units to set $\lambda = 1$.
	
	The numerical implementation of the flow equations is available on \href{https://github.com/laurabatini/flow-equations-code}{Github}~\cite{Githubcode}.

	\section{Static and dynamic criticality}
	\label{sec:crit_dynamics}
	
	Having introduced the field-theory representation of the relaxation dynamics model and presented its flow equations in Sec.~\ref{sec:CriticalPhenomena}, we analyze the second-order phase transition of this theory.
	In the vicinity of the phase transition, the system becomes strongly coupled, and the correlation length, denoted as $\xi$, diverges as we tune toward criticality
	\begin{equation}
		\xi(t_r) = \xi_{\pm} |t_r|^{-\nu}
		\, ,
	\end{equation}
	where $t_r$ is the reduced temperature, defined as
	\begin{equation}
		t_r \equiv \frac{m^2-m_c^2}{|m_c|^2}
		\, ,
		\label{eq:redtemp}
	\end{equation}
	where $m_c$ is the critical mass.
	
	\begin{figure*}[t]
		\centering
		\captionsetup{justification=raggedright,singlelinecheck=false}
		\begin{subfigure}[t]{0.48\textwidth}
			\centering
			\includegraphics[width=\linewidth]{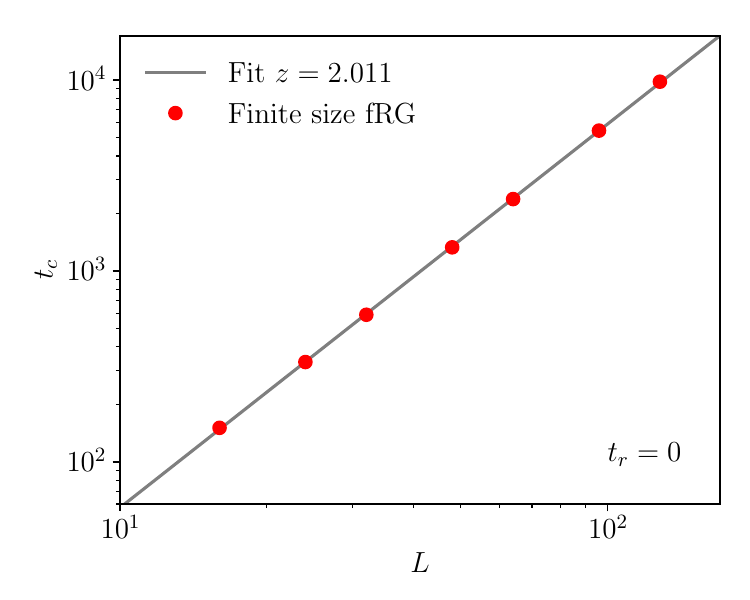}
			\label{fig:Finite_size_z}
		\end{subfigure}%
		\begin{subfigure}[t]{0.48\textwidth}
			\centering
			\includegraphics[width=\linewidth]{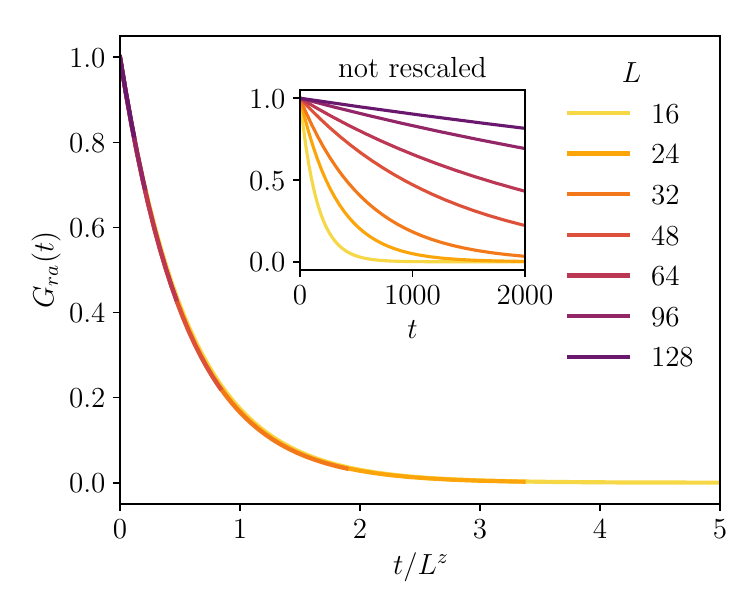}
			
			\label{fig:G_n_rescaled}
		\end{subfigure}	~
		
		\caption{\textbf{Left:} Relaxation time $t_c(0, L) $ cf. \labelcref{eq:rescaled} versus linear box size $L$ computed using the finite size RG flow at the critical mass $m^2=m_c^2$. The solid line shows the best fit for the dynamic critical exponent $z= 2.011$. \textbf{Right:} Retarded correlator at vanishing spatial momentum $G_{ra}(t, \boldsymbol{p}=0)$ as a function of rescaled time $t/L^z$ for different box lengths $L$. The curves become independent from the size after rescaling the time by $L^{-z}$. The inset displays the retarded correlator without any rescaling applied. }
		\label{fig:please_change_me_before_use_2}
	\end{figure*}
	
	As the correlation length increases, the characteristic relaxation time $t_c$ diverges at the phase transition point, leading to the phenomenon known as critical slowing down. This behavior is typically associated with the correlation length, $\xi$, through the dynamic critical exponent, $z$, according to the relation (for further details, see, e.g.,~ \cite{Zinn-Justin:572813, Tauber:2016mpa}):
	\begin{equation}
		t_c(t_r) \sim \xi^{z} (t_r)\sim |t_r|^{-z \nu}
		\, .
	\end{equation}
	
	The setting presented in Sec.~\ref{sec:CriticalPhenomena} provides us with the tools to investigate static and dynamical criticality simultaneously.
	In this section, we present numerical results for the critical behavior of the theory.
	This serves primarily as a benchmark of our setup, as they have been studied previously within the fRG formalism~\cite{Canet:2002gs, Canet:2003qd, DePolsi:2021cmi, Murgana:2023xrq}.
	

	\subsection{Static	critical exponents} \label{subsec:Static critical exponents}
\begin{figure*}[t]
	\centering
	\captionsetup{justification=raggedright,singlelinecheck=false}
	\begin{subfigure}[t]{0.48\textwidth}
		\centering
		\includegraphics[width=\linewidth]{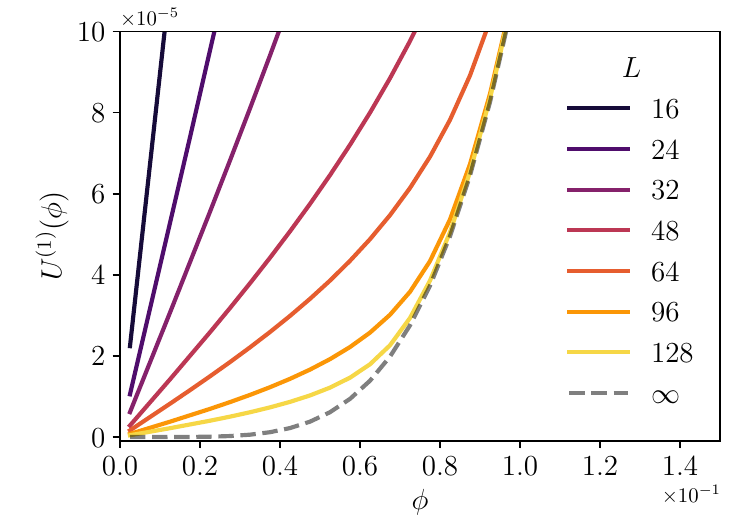}
		\label{FinitesizeU}
	\end{subfigure}%
	\begin{subfigure}[t]{0.48\textwidth}
		\centering
		\includegraphics[width=\linewidth]{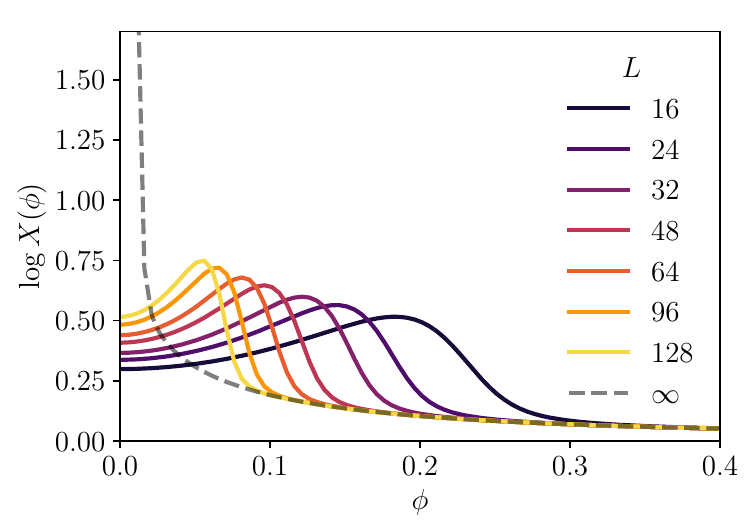}
		\label{FinitesizeX}
	\end{subfigure}	~
	
	\caption{ Derivative of the effective potential $U^{(1)}(\phi)$ and logarithm of the dissipation rate $\log X(\phi)$ at the critical mass $m^2 \approx m^2_c$  for various box lengths $L$.
		The dashed line represents the infinite-volume limit. 
		\textbf{Left:} Derivative of the effective potential $U^{(1)}(\phi)$ as function of the field $\phi$. Increasing the size $L$, the derivative of the effective potential becomes flatter in the vicinity of the minimum in $\phi=0$. The dashed line represents the infinite-volume limit $L \to \infty$, which is flat in the origin. \textbf{Right:} Logarithm of the dissipation rate $\log X(\phi)$ as function of the field $\phi$. Increasing the size $L$, the height of the barrier increases and shifts towards $\phi=0$. In the infinite-volume limit, the barrier diverges at the origin.}
	\label{fig:finitesizecriticalmass}
\end{figure*}
	
	The static scaling relation for the order parameter, given by the expectation value of the field $\phi_0 \equiv \langle \phi \rangle$, is described by the following behavior:
	\begin{equation}
		\label{eq:beta}
		\phi_0(t_r)=a \Theta(t_r) |t_r|^\beta,
	\end{equation}
	where $\beta$ is the critical exponent for the order parameter.
	
	Similarly, to compute $\nu$, we exploit the fact that  the correlation length $\xi$ is inversely proportional to  the renormalized mass near the phase transition 
	\begin{equation}
		\xi \sim m_R^{-1} \sim  t_r^{-\nu}
		\, .
	\end{equation}
	In the symmetric phase, the square of the renormalized mass is given by
	\begin{equation}
		m_R^2= \lim_{k \to 0} U^{(1)}_{k}(\phi_0)
		\, .
	\end{equation}
	
	We access the phase transition by scanning the bare mass parameter $m$, which is the only free parameter at hand.
	We solve the RG flow deep into the IR, where we determine the expectation value of the field $\phi_0$.
	Our final RG scale is chosen as
	\begin{equation}
		k_\tinytext{IR} = \Lambda e^{-9.5} \approx 0.0007485
		\, ,
	\end{equation}
	which corresponds to an RG time of $9.5$.
	We then extract the critical scaling by performing ordinary $\chi^2$ minimization.
	The expectation value of the field is shown as a function of the reduced temperature in the left panel of Fig.~\ref{fig:Critical_fits}.
	By performing an ordinary $\chi^2$ minimization on the scaling law \labelcref{eq:beta}, we find the critical bare mass parameter $m_c^2= -0.927814$ and the exponent $\beta = 0.36$.
In the fit (the same reasoning applies to the discussion of $\nu$ below), we omit the data point closest to the critical mass due to the breakdown of our approximation near this critical point.
	Note that the mass parameter has to be determined very precisely since, in these units, the scaling window is comparably small; cf.~Fig.~\ref{fig:Critical_fits}, and e.g., \cite{Schaefer:2006ds}.
	
	Our result for the exponent $\beta$ agrees with the well-known static exponents for the three-dimensional Ising universality class.
	Previous fRG studies~\cite{Canet:2002gs, Canet:2003qd, DePolsi:2021cmi} found $\beta = 0.3486(59)$ in the local potential approximation (LPA) and $\beta = 0.3263(4)$ in the fourth-order derivative expansion.
	State-of-the-art results from other methods include $\beta = 0.32643(6)$ from Monte Carlo methods~\cite{Hasenbusch_2010} and $\beta = 0.32599(32)$ from an $\epsilon$ expansion~\cite{Kompaniets:2017yct}.
	
	The correlation length as a function of the reduced temperature is shown in the right panel of Fig.~\ref{fig:Critical_fits}.
	Proceeding as for the exponent $\beta$, we find $\nu = 0.67$.
	This is, again, in decent agreement with existing results:
	fRG studies~\cite{Canet:2002gs, Canet:2003qd, DePolsi:2021cmi} found $\nu = 0.634(8)$ in the LPA and $\nu = 0.62989(25)$ in the fourth-order derivative expansion.
	The Monte Carlo estimate~\cite{Hasenbusch_2010} is $\nu = 0.63002(10)$ and the $\epsilon$ expansion estimate~\cite{Kompaniets:2017yct} is $\nu = 0.6292(5)$.
	
	Note that the exponents $\nu$ and $\beta$ are not independent without higher-order terms in gradients, such as a renormalization of the spatial derivative term in the effective action (in our case, the anomalous critical exponent is $\eta=0$).
	Thus, the critical exponents obey the scaling relation  $2\beta=\left(1+\eta\right) \nu = \nu$.
	This could be used to give a rough estimate of the error.
	The error of our calculation is influenced by several factors, two of the most prominent ones being finite numerical precision and fit errors. Additionally, as one approaches criticality more closely, the system becomes increasingly sensitive to deviations from the critical value. 
	Using more sophisticated truncations to higher orders in the derivative expansion would be required to obtain more precise results, which goes beyond this work. Given that this aspect of our study serves primarily as a benchmark, and our exponents align qualitatively and quantitatively with existing literature, we refrain from estimating the error for our results in detail.

	\subsection{Dynamic critical exponent} \label{subsec:Dynamic critical exponents}

	In addition to the static properties, the dynamic universality class is further characterized in terms of the scaling exponent $z$.
	The retarded correlation function~$G_{ra}( t ,\boldsymbol{x})= \langle \phi_r(t, \boldsymbol{x}) \phi_a(0,0)\rangle$ has the following scaling form in Fourier space~\cite{tauber2014}
	\begin{equation}
		\label{eq:scaling_tpf}
		G_{ra}(\omega,\boldsymbol{p}) = \frac{1}{|\boldsymbol{p}|^{2-\eta}} {\chi}(\omega \xi^{z}, |\boldsymbol{p}|\xi) 
		\,,
	\end{equation}
	where $\chi$ is an analytic universal function.
	This relation can be used to define the dynamic critical exponent $z$. 
	In our approximation, where the effective action takes the form \labelcref{FieldtheoryEffaction} with  the Hamiltonian \labelcref{eq:Isinghamiltonian}, the retarded function takes the form (cf. Appendix~\ref{app:dyn_field_theory}),
	\begin{align}
		G_{ra}(\omega, \boldsymbol{p})=\frac{1}{-i  X \omega+U^{(2)}+ \boldsymbol{p}^2}
		\, ,
	\end{align}
	which can be cast into the scaling form by extracting the trivial momentum dependence,
	\begin{align}
		G_{ra}(\omega,\boldsymbol{p})=\frac{1}{\boldsymbol{p}^2 }\frac{\boldsymbol{p}^2/U^{(2)}}{-i X \omega/U^{(2)}+1 +\boldsymbol{p}^2/U^{(2)}}
		\, .
	\end{align}
	Now, by defining the correlation length $\xi$ and relaxation time $t_c$ as
	\begin{equation}
		\label{t_c}
		\xi(t_r) \equiv \sqrt{ \frac{1}{U^{(2)}}} \quad  \text{and} \quad t_c(t_r) \equiv \frac{X}{U^{(2)}}
		\, ,
	\end{equation} 
	the scaling relation \labelcref{eq:scaling_tpf} takes the form
	\begin{align}
		G_{ra}(\omega,\boldsymbol{p})=\frac{1}{\boldsymbol{p}^2 }\chi(\omega t_c,|\boldsymbol{p}| \xi)
		\, .
	\end{align}
	In real space, i.e.,~after a Fourier transform, the retarded function is given by~\cite{Zinn-Justin:572813}
	\begin{equation}
		\begin{aligned}
			G_{ra}(t,\boldsymbol{p})&=\int \frac{\mathrm{d} \omega}{2 \pi} \frac{1}{-i X  \omega+U^{(2)}+\boldsymbol{p}^2} \exp (-i \omega t) \\& =\exp \Big(-\frac{U^{(2)}+\boldsymbol{p}^2}{X} t \Big) \Theta(t)
			\, ,
		\end{aligned}
		\label{Gra}
	\end{equation}
	where $\Theta$ is the Heaviside step function.

	The dynamic critical exponent can now conveniently be extracted with a finite-size scaling analysis.
	In fact, the size of the system provides an additional length scale in the system. The scaling relation of the relaxation time in a finite volume of linear size $L$ is given by	 \cite{Zinn-Justin:572813}
	\begin{equation}
		\label{eq:rescaled}
		t_c(t_r, L)= L^z f_{t_c} (t_r L^{1 / \nu})
		\, .
	\end{equation}
	
	We work sufficiently close to the critical point to approximate $f_{t_c} \approx f(0)$.
	In this case, we confine the system in a three-dimensional box of linear size $L$, cf.~Appendix~\ref{app:flows} for details on the modifications to the flow equations in a finite volume.
	In the left panel of Fig.~\ref{fig:please_change_me_before_use_2}, we show the relaxation time $t_c$ as a function of the box length $L$ at the critical temperature, i.e.,~$t_r = 0$. We obtain the fit $z = 2.011$, in good agreement with theoretical predictions using the $\epsilon$ expansion $z \approx 2.02 $ \cite{Zinn-Justin:572813, kamenev_2011} and  from Monte Carlo simulations where $z =2.026(56)$ \cite{Schaefer:2022bfm}.

	The correlation function \labelcref{Gra} collapses by rescaling the time $t$ by $L^z$, as shown in the right panel of Fig.~\ref{fig:please_change_me_before_use_2}, where $z$ is the result of the fit above.
	
	\subsection{Field dependence at criticality}
	
	Finally, it is insightful to gain a qualitative understanding of the field dependence of $U^{(1)}(\phi)$ and $X(\phi)$ close to criticality. This applies to both infinite and finite volumes.
	In the left panel of Fig.~\ref{fig:finitesizecriticalmass}, the derivative of the effective potential is shown for different finite volumes. As the volume increases, the derivative near the origin, representing the mass term, gradually diminishes as expected at the critical point in the limit of infinite-volume (indicated by the dashed line).
	
	For the more intriguing case, the dissipation rate, we present its logarithm in the right panel of Fig.~\ref{fig:finitesizecriticalmass}.
	In this instance, remarkably, a barrier forms gradually, exhibiting divergence in the infinite-volume limit at the origin.
	A direct consequence thereof can be inferred from \labelcref{Gra}. At criticality, the mass term $U^{(2)}$ vanishes, leading to an infinite correlation length at zero spatial momentum, i.e.,~critical slowing down.
	However, due to the divergence of the dissipation rate, critical slowing down is observed for all spatial momenta.
	
	\section{Results in the ordered phase}
	\label{sec:non_crit_results}
	
	\begin{figure}[t]
		\centering
		\includegraphics[width=.4\textwidth]{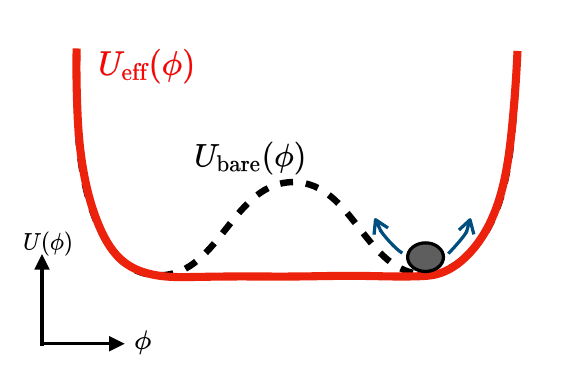}
		\caption{Sketch of the symmetry-breaking potential $U(\phi)$. The bare potential $U_{\mathrm{bare}}= U_{\Lambda}$ has two degenerate minima, where the field sits and fluctuates around, separated by a potential barrier. The corresponding effective potential $U_{\mathrm{eff}} = U_{k\to0}$, which incorporates all the fluctuations, is flat. }
		\label{fig:Symmetry breaking}
	\end{figure} 
	
	\begin{figure*}[t]
		\centering
		\captionsetup{justification=raggedright,singlelinecheck=false}
		\begin{subfigure}[t]{0.48\textwidth}
			\captionsetup{justification=raggedright,singlelinecheck=false}
			\includegraphics[width=\linewidth]{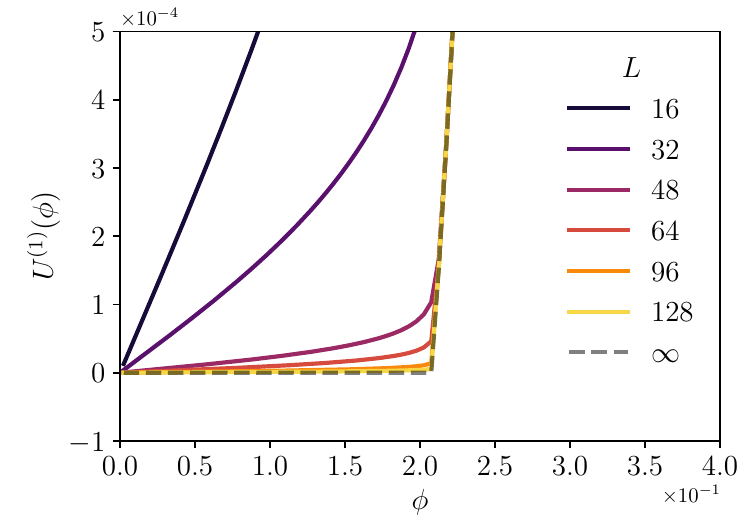}
		\end{subfigure}%
		\begin{subfigure}[t]{0.48\textwidth}
			\includegraphics[width=\linewidth]{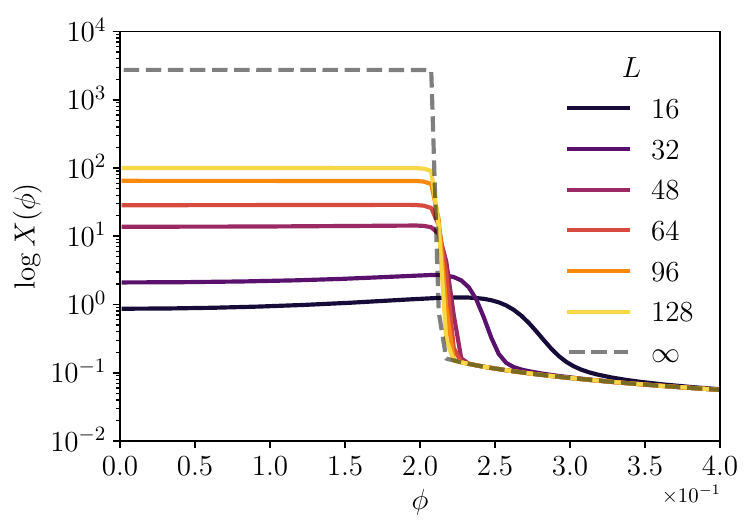}
		\end{subfigure}	~
		\caption{Derivative of the effective potential $U^{(1)}(\phi)$ and logarithm of the dissipation rate $X(\phi)$ at the initial squared mass $m^2 = -0.9450$ (broken phase)  displayed for different system sizes.
			The dashed line represents the infinite-volume limit. 
			\textbf{Left:} 	Derivative of the effective potential $U^{(1)}(\phi)$ as a function of the field $\phi$.	With increasing size $L$, the derivative of the effective potential becomes flatter near the potential minimum. \textbf{Right:} Logarithm of the dissipation rate $\log X(\phi)$ as a function of the field $\phi$. With increasing size $L$, the height of the barrier increases.}
		\label{fig:Uprime_and_X_finiteL}
	\end{figure*}

In the previous section, our focus was on the critical regime. Now, we shift our attention to the ordered phase characterized by spontaneously broken symmetry, i.e., $t_r < 0$.
The example of spontaneous symmetry breaking in a $\phi^4$ theory is pivotal in our current understanding of spontaneous symmetry breaking. The dynamics of the order parameter potential exemplify the behavior observed in virtually all second-order phase transitions, thus underscoring its paramount importance.

As discussed in Sec.~\ref{sec:Hydro}, order parameter potentials of this nature often come into play in effective descriptions. In these descriptions, the full effective potential vanishes for field values below the minimum. A visual representation of this is depicted in Fig.~\ref{fig:Symmetry breaking}. However, it is noteworthy that the input potentials employed in nearly all cases closely resemble the bare potential. This prompts the question of how to reconcile the full effective potentials, which arise as the proper results in quantum field theory calculations, with the input used in effective theories.

This problem is less pronounced within RG flows because convexity is only restored in the limit $k\to0$ when all fluctuations are integrated out. Hereby, the dominant contribution at the solution of the equation of motion is generated at a finite RG scale, where the potential still resembles a double-well.

A typical example of an effective potential in the broken phase is shown in the left panel of Fig.~\ref{fig:Uprime_and_X_finiteL}. Additionally, looking at different finite volumes in parallel is instructive again. As required, the potential is convex in all cases but only in the infinite-volume case is it truly flat. For the following discussion, it is noteworthy that the potential is not exactly zero in the infinite-volume case. This is due to the finite final RG scale necessary in numerical applications. In principle, it would be possible to extrapolate the resulting potential to zero (see e.g.~\cite{Grossi:2019urj}), but all conclusions can be drawn here without doing so.

The dissipation rate gives the answer to our question regarding the field dynamics in this domain, shown in the right panel of Fig.~\ref{fig:Uprime_and_X_finiteL}. A barrier emerges that grows exponentially with volume. Its value for infinite volume is only finite due to the finiteness above the final RG scale.
Consequently, plugging the combination of a flat potential with a diverging dissipation into dynamic evolution equations such as \labelcref{josephson-diss} shows that the field can never enter the flat regime of the potential.

Since the transition appears as a rigid boundary, dynamically, the field will most likely be reflected and bounce back when approaching the minimum of the potential.
However, at the minimum of the potential, $U^{(1)}(\phi_0)=0$ by definition, the potential is symmetric to the leading order at the minimum. This justifies the use of effective double-well potentials at leading order, although it should be mentioned that it cannot capture higher-order effects adequately.

This situation differs from the critical scenario described in Sec.~\ref{sec:crit_dynamics}. There, the boundary scales towards the minimum at $\phi_0 = 0$, while in the broken phase, a rigid boundary forms at a nonvanishing field expectation value $\phi_0 \neq 0$.

\begin{figure*}[t]
	\centering
	\captionsetup{justification=raggedright,singlelinecheck=false}
	\begin{subfigure}[t]{0.48\textwidth}
		\captionsetup{justification=raggedright,singlelinecheck=false}
		\includegraphics[width=\linewidth]{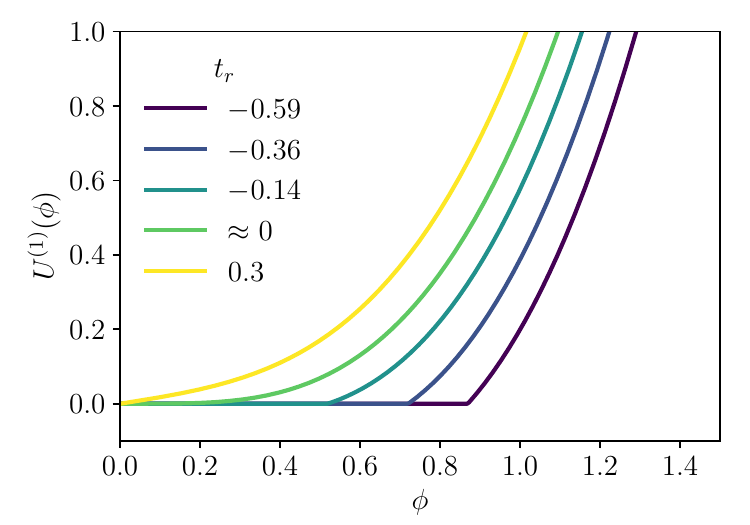}
		\label{Uprime_masses}
	\end{subfigure}%
	\begin{subfigure}[t]{0.48\textwidth}
		\captionsetup{justification=raggedright,singlelinecheck=false}
		\includegraphics[width=\linewidth]{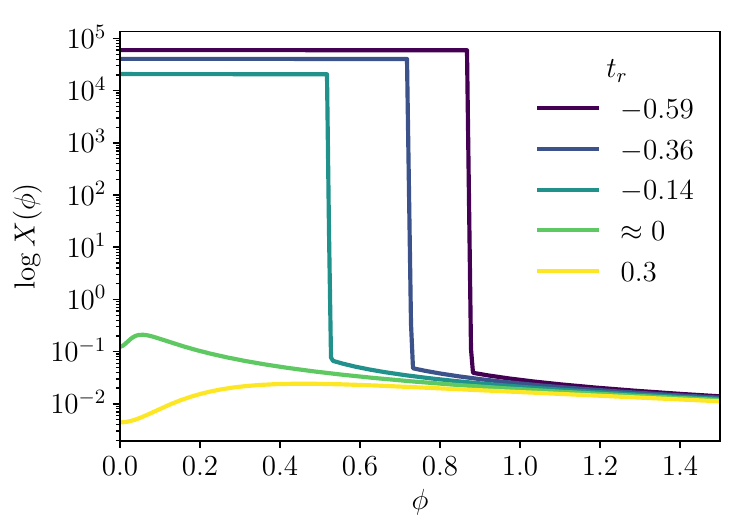}
	\label{fig:tau_infinite}
\end{subfigure}	~
\caption{Derivative of the effective potential $U^{(1)}(\phi)$ and logarithm of the dissipation rate for various reduced masses $t_r$.
	\textbf{Left:} Derivative of the effective potential $U^{(1)}(\phi)$ as a function of the field $\phi$. When $t_r<0$, the derivative of the effective potential exhibits a flat region between its minima. \textbf{Right:} Logarithm of the dissipation rate $\log X(\phi)$ as a function of the field $\phi$. In the case of $t_r<0$, a diverging barrier emerges within the same field range where the potential undergoes flattening. }
\label{fig:Uptau_phase_transition}
\end{figure*}

The transition between different regimes of the theory is depicted in Fig.~\ref{fig:Uptau_phase_transition}. While the derivative of the potential changes in a continuous manner, the changes in the dissipation rate are much more abrupt.
Its scaling at criticality is not fully visible due to a slight numerical detuning from criticality and the enormous size of the dissipation rate in the broken phase, which is only restricted by the final RG scale. For a proper visualization of the scaling at criticality, see the right panel of Fig.~\ref{fig:finitesizecriticalmass}.

The general behavior of the dissipation rate with respect to a phase transition, i.e., its divergence in the flat part, is generic due to the general nature of the underlying flow equations.
The fact that it is almost constant in the physical regime of the potential and could hence be approximated by a constant may be solely dependent on the specific model being considered.

\section{Conclusions} \label{sec:Conclusion}
In this study, we used the fRG formulated on the Schwinger-Keldysh contour to analyze the relaxation dynamics of a simple $\phi^4$ theory with dissipation, which reduces to model A at its second-order phase transition. Using a derivative expansion up to first order in time and second order in space, we retained the full field dependence of the effective potential and the dissipation rate.
To validate our approach, we first determined the static critical exponents of model A.
We confined our system to a periodic box to extract the dynamic critical exponent and performed a scaling analysis.

Furthermore, we investigated the field dependence of the dissipation rate in the ordered phase of the theory, away from criticality.
It diverges in the regime of field space where the effective potential is flat, i.e., for expectation values smaller than the physical one.
This result has implications for using full effective potentials, which must be convex, in effective long-range theories, such as hydrodynamic simulations.
The divergence of the dissipation rate implies that this regime of the theory is not accessible in effective nonequilibrium descriptions in the linear response regime and is screened by a rigid barrier.
As an outlook, we want to extend our analysis to explore other dynamic universality classes beyond model A.
One of our key objectives is to delve into the critical dynamics of model G, which corresponds to two-flavor QCD in the chiral limit. This broader exploration is motivated by the potential relevance for heavy-ion collision phenomenology.

\section{Acknowledgements} \label{Acknowledgments}
We thank Jürgen Berges, Sebastian Diehl, Johannes Roth and Pierfrancesco Urbani for the stimulating discussions.
E.G. would like to thank Stefan Flörchinger, Derek Teneay and Adrian Florio for discussions and work on related topics.
L.B. acknowledges support by the Deutsche Forschungsgemeinschaft (DFG, German Research Foundation) under the Collaborative Research Center SFB 1225 ISOQUANT (Project-ID 27381115).
N.W. acknowledges support by the Deutsche Forschungsgemeinschaft (DFG, German Research Foundation) – Project number 315477589 – TRR 211 and by the State of Hesse within the Research Cluster ELEMENTS (Project No. 500/10.006).

\bookmarksetup{startatroot}
\appendix



\section{Dynamical field theory} \label{app:dyn_field_theory}

For the reader's convenience, we provide here a short introduction to the Schwinger-Keldysh functional integral formalism \cite{Keldysh:1964ud}, which provides a framework for the field-theoretical description of real-time dynamics. 
This formalism can be applied to the field-theoretical description of the hydrodynamic (coarse-grained) system as described in Sec.~\ref{sec:Hydro} 
(see~\cite{Liu:2018kfw} for details).

Let us consider the functional integral representation of the so-called Schwinger-Keldysh partition function: $\mathcal{Z}_\beta=\operatorname{tr}\left(e^{-\beta \mathcal{H}}\right)$. The Hamiltonian $\mathcal{H}$ generates the unitary dynamics.

By introducing different sources $J_{\pm}$ for the fields on the two branches, the partition function is
\begin{equation}
	\mathcal{Z}[J_+, J_-] = \tr \Big( U^{\dagger}_{J_-}(-\infty, +\infty) e^{- \beta \mathcal{H}} U_{J_+}(-\infty, +\infty)\Big)/\mathcal{Z}_\beta \, ,
	\label{eq:Zpm}
\end{equation}
where $U_{J_{\pm}}$ is the unitary evolution operator with external source $J_{\pm}$ coupled to the field.
The functional derivatives of \labelcref{eq:Zpm} can be used to generate time-dependent correlation functions.
In this expression, time evolution can be interpreted as occurring along a closed path with two branches, the forward and the backward branch of the time path, producing a closed-time path-integral. It is convenient to introduce $\varphi_+(t,\boldsymbol{x})$ and $\varphi_-(t,\boldsymbol{x})$ fields, where the subscripts $\pm$ indicate that the respective time arguments of the sources are taken on the forward or backward branch of the closed time contour, respectively.
The partition function in the path integral representation can be written as
\begin{equation}
	\mathcal{Z}[J_+, J_-] = \int \mathcal{D} \varphi_+  \mathcal{D} \varphi_- \rho[\varphi_+, \varphi_-]e^{i S[\varphi_+, J_+]- iS[\varphi_-, J_-]} \, ,
\end{equation}
where we denote the microscopic action of the field as $S[\varphi_{\pm}, J_{\pm}]= S[\varphi_{\pm}] + i \int_x J_{\pm} \varphi_{\pm}$, $\int_x \equiv \int_{-\infty}^{\infty} \mathrm{d} t \int \mathrm{d}^d x \equiv \int \mathrm{d}^{d+1} x$, and the density matrix $\rho$ is a functional of $\varphi_+$ and $\varphi_-$ at the initial time $t\to -\infty$.

It is useful to move from the $\pm$ basis to the so-called $r$-$a$ basis, where the fields and external current are defined as the symmetric and antisymmetric combinations of fields on the forward and backward branches, respectively.
The components are called classical $(r)$ and quantum $(a)$.
Explicitly, the components are given by
\begin{equation} 
	\label{eq:Keldysh_basis}
	\begin{aligned}
		&\varphi_r(t,\boldsymbol{x})=\frac12( \varphi_+(t,\boldsymbol{x}) + \varphi_-(t,\boldsymbol{x})) \, ,
		\\ &\varphi_a(t,\boldsymbol{x}) = \varphi_+(t,\boldsymbol{x}) - \varphi_-(t,\boldsymbol{x})	 
		\, ,
	\end{aligned}
\end{equation}
and 
\begin{equation} 
	\label{eq:Keldysh_basis}
	\begin{aligned}
		J_r(t, \boldsymbol{x})&=\frac12( J_+(t,\boldsymbol{x}) + J_-(t,\boldsymbol{x})) \,,
		\\ 	J_a(t, \boldsymbol{x})&= J_+(t,\boldsymbol{x})-J_-(t,\boldsymbol{x})
		\, .
	\end{aligned}
\end{equation}
We are going to use a matrix notation and define the following vectors
\begin{equation}
	\varphi(t,\boldsymbol{x})\equiv \left(\begin{array}{c}
		\varphi_r(t,\boldsymbol{x}) \\
		{\varphi_a}(t,\boldsymbol{x})
	\end{array}\right) \quad \text { and } \quad J(t,\boldsymbol{x}) \equiv \left(\begin{array}{c}
		J_r(t,\boldsymbol{x}) \\
		{J_a}(t,\boldsymbol{x})
	\end{array}\right) \, .
\end{equation}
The generic consequence of the double path formulation is that if the currents are set to equal,
\begin{equation}
	\mathcal{Z}[J_+ = J,J_-=J ]= 1 
	\, ,
\end{equation} and in the $r$-$a$ formulation $\mathcal{Z}[J_r = J, J_a = 0 ] = 1$.
Taking functional derivatives with respect to the $J_r$ current generates the $n$-point connected correlation functions, 
\begin{equation}
	\label{eq:unitarity}
	(-i)^{n+1} 	\frac{\delta^n  \log \mathcal{Z}[J_r, J_a]}{\delta J_r \cdots \delta J_r} = G_{a \cdots a} = 0 
	\, .
\end{equation}
A system is in thermodynamic equilibrium if and only if the corresponding Schwinger-Keldysh action is invariant under the transformation $\mathcal{T}_\beta$, as defined in \cite{Sieberer:2015hba, sieberer2016keldysh} for scalar fields:
\begin{equation}
	\mathcal{T}_\beta \varphi_{\pm}(t,\boldsymbol{x})=\varphi_{\pm}(-t \pm i \beta / 2,\boldsymbol{x}) =e^{ \pm i \frac{\beta}{2} \partial_t} \varphi_{ \pm}(-t, \boldsymbol{x})
	\, .
\end{equation}
This transformation is simpler to realize in the semiclassical limit, where $e^{ \pm i \frac{\beta}{2} \partial_t} \approx 1 \pm i \frac{\beta}{2} \partial_t$. The action on the fields is explicitly given in the $r$-$a$ basis by
\begin{equation}
	\begin{aligned}
		\mathcal{T}_\beta \varphi_r(x)  &=    \,\Theta \varphi_r(x)=\varphi_r\left(t_R, \boldsymbol{x}\right)   \, , \\
		\mathcal{T}_\beta \varphi_a(x) & \simeq  \, \Theta \varphi_a(x)+i \Theta \partial_t \varphi_r(x) \\ & = \varphi_a\left(t_R, \boldsymbol{x}\right)-i \partial_{t_R} \varphi_r\left(t_R, \boldsymbol{x}\right) 	\, ,
	\end{aligned}
	\label{Symmetry}
\end{equation}
where the action of $\Theta$ is a time reversal $\Theta t = -t \equiv t_R$.
This dynamical symmetry implies the Kubo-Martin-Schwinger (KMS) condition \cite{Kubo:1957mj, PhysRev.115.1342}. 
Furthermore, the equilibrium two-point correlation functions are required to satisfy a detailed balance condition, equivalent to the fluctuation-dissipation relation. 

We are interested in the effective action $\Gamma[\Phi]$, defined by the Legendre transform of the Schwinger functional $W = -i \log \mathcal{Z}\left[J_r, J_a\right]$, i.e., 
\begin{equation}
	\Gamma\left[\Phi\right]=\sup _{J_a, J_r}\left\{W-	\int_{t,\boldsymbol{x}}\left(J_a \phi_r+J_r \phi_a \right)\right\} , 
\end{equation}
in terms of the field expectation values $\Phi = \langle \varphi \rangle$.  
Unitarity, i.e., \labelcref{eq:unitarity}, leads to the fact that all monomials of the field $\phi_r$ are zero in the effective action \cite{Liu:2018kfw}.
Given the KMS symmetry and unitarity, a consistent ansatz for the effective action can be organized in powers of temporal derivatives in the semiclassical approximation as
\begin{equation}
	\Gamma[\Phi] = \Gamma_1[\Phi] + \Gamma_2[\Phi] + \dots \, .
\end{equation}
Further symmetries are parity and the $\mathcal{Z}_2$ invariance in the fields. 

In the following two subsections, we show explicitly how to formulate the constraints on the allowed terms of the action in first and second order in temporal derivatives.
\subsection{Truncation at first order in time derivatives}
At first order in time derivatives, the action  can be parametrized as 
\begin{equation}
	\begin{aligned}
		\Gamma[ \Phi]   = \int_{t,\boldsymbol{x}} & \Big\{ \phi_aF(\phi_r, \nabla \phi_r) +X(\phi_r, \nabla \phi_r)\phi_a\partial_t \phi_r 
		\\ &
		+ H(\phi_r, \nabla \phi_r) \phi_a^2 \Big\}
		\, .
	\end{aligned}
\end{equation} 
As stated above, we have two more symmetries, $\mathcal{Z}_2$, and parity. Therefore, in the previous equation, we have further constraints on the functions $F, X$, and $H$.
$\mathcal{Z}_2$ symmetry implies 
\begin{equation}
	\label{eq:Z_2}
	F \to -F, \quad X \to X, \quad  H \to H  \,.
\end{equation}
Moreover, $F$ depends on odd powers of $\phi_r$, while $X, H$ depend on even powers, and all those functions must be invariant under parity transformation.

The consequence of the KMS symmetry reduces the number of allowed terms. 
Applying the transformation in \labelcref{Symmetry}, the action transforms to 
\begin{equation}
	\begin{aligned}	  
		\Gamma[ \mathcal{T}_\beta \Phi] =	\int_{t_R,\boldsymbol{x}}  &\Big\{ \phi_aF +i   F \partial_{t_R} \phi_r +  (-X 	+ 2 i  \beta H)\phi_a\partial_{t_R} \phi_r  \\ & 
		+ H \phi_a^2-	(	\beta^2 H +  i \beta X )(\partial_{t_R} \phi_r)^2 
		\Big \}  \, .
	\end{aligned}
\end{equation}
The action is invariant if $H= - i X/ \beta$ and if $F=\frac{\delta \mathcal{H} [ \phi_r] } {\delta \phi_r}$, such as when choosing as $\mathcal{H}$ the Ising Hamiltonian \labelcref{eq:Isinghamiltonian}, with the quartic potential \labelcref{eq:phi4potential}.
The second requirement is needed to make the monomial $i F \partial_t \phi_r$ a total derivative,
\begin{equation}
	i \frac{\delta \mathcal{H} [ \phi_r ]}  {\delta \phi_r} \partial_t \phi_r = i \partial_t h ( \phi_r ) 
	\, , 
\end{equation} 
where 
\begin{eqnarray}
	\mathcal{H}[\phi_r] = \int_{t, \boldsymbol{x}} h(\phi_r, \nabla \phi_r)  \, , 
\end{eqnarray}
such that unitarity is expected. 
This term can be interpreted as the system's entropy \cite{Glorioso:2016gsa}.

To summarize, we have the following consistent ansatz for the effective action 
\begin{equation}
	\label{FieldtheoryEffaction}
	\Gamma [\Phi] = 	\int_{t,\boldsymbol{x}} \Big\{ \phi_a\frac{\delta \mathcal{H}}{\delta \phi_r}+X \Big(\phi_a\partial_t \phi_r -\frac{i}{\beta} \phi_a^2 \Big) \Big\}
	\, ,
\end{equation}
where we notice that the time derivative and the noise term have the same coefficient $X$.
After we set without loss of generality $\beta = 1$, the resulting effective action has the form of the response functional of the equilibrium dynamical model A.
The corresponding equation of motion for $\phi_r$ is nothing but the Langevin evolution, given in \labelcref{eq:langevin_evolution}.

We also state the propagator, which derives from~\labelcref{FieldtheoryEffaction} for completeness and later convenience.
For additional context see~\labelcref{propagator}.
In Fourier space, it is given in matrix form as
\begin{eqnarray}
	G(\omega,\boldsymbol{p})=\frac{1 }{\gamma\left(p\right) \gamma\left(-p\right)}\left(\begin{array}{cc}
		-2 i X & \gamma\left(-p\right) \\
		\gamma\left(p\right) & 0
	\end{array}\right),
\end{eqnarray}
with 
$\gamma\left(p\right) \equiv U^{(2)}+\boldsymbol{p}^2-i X \omega$.
\subsection{Truncation at second order in time derivatives}
It is possible to extend the truncation one order higher in time derivatives using the algorithm explained in \cite{Glorioso:2016gsa}. 

Within the semiclassical expansion, we can write the ansatz for the action up to the third order in $\phi_a$ and two time derivatives 
\begin{equation}
	\label{eq:tuma}
	\begin{aligned}
		\Gamma[\Phi] =
		\int_{t,\boldsymbol{x}} \mathcal{L} = \int_{t,\boldsymbol{x}} &\Big \{  
		E \phi_a  + X\phi_a^2  +Y \phi_a^3\Big\} \, ,
	\end{aligned}
\end{equation}
where $E, X, Y $ are functions of the $r$-field $\phi_r$ and its time and spatial derivatives are consistent with the $\mathcal{Z}_2$ symmetry of the Ising model.
The KMS symmetry dictates the number of time derivatives in each coefficient contributing to the same order; if we include the cubic term in $\phi_a$, to close the KMS symmetry, we have to expand to the second order in the time derivative of the coefficients  in \labelcref{eq:tuma}
\begin{equation}
	\begin{aligned}
		E & = E_0 + E_1 + E_ 2\, , \\
		X &= X_0 + X_1 \, , \\
		Y &= Y_0 \, .
	\end{aligned}
\end{equation}
The subscript counts the order in time derivatives of the $\phi_r$ terms. 
The action of the transformation on $\mathcal{L}$ is 
\begin{equation}
	\begin{aligned}
		\mathcal{L}[ \mathcal{T}_\beta \Phi]  &  = 
		i \beta   E \partial_t \phi_r - \beta^2  (\partial_t \phi_r)^2 X - i \beta^2 (\partial_t \phi_r)^3 Y 
		\\ & 
		+  \Big( 
		E+ 2i \beta  X\partial_t \phi_r  + 3 (i \beta \partial_t \phi_r)^2 Y  \Big)\phi_a 
		\\ & 
		+ \Big( X + 3 i \beta \partial_t \phi_r Y  \Big) \phi_a^2 
		\\ & 
		+ \phi_a^3 Y \, . 
	\end{aligned}
\end{equation}
For the action to be invariant $\Gamma[\Phi] =  \Gamma[ \mathcal{T}_{\beta} \Phi ]+ \int_{t, \boldsymbol{x}} \partial_{\mu} V^{\mu}$. 
We get the following relations for the coefficients.
At order zero in powers of the $a$ field, we get the following conditions:
\begin{equation}
	\begin{aligned}
		& i \beta  E_0 \partial_t \phi_r  = \partial_\mu V^{\mu}_{(0,0)}, \\ & 
		i\beta E_1 \partial_t \phi_r - \beta^2 X_0 (\partial_t \phi_r)^2 = \partial_\mu V^{\mu}_{(0,1)},  \\ &
		i\beta E_2  \partial_t \phi_r - \beta^2 X_1(\partial_t \phi_r)^2
		- i \beta^3 (\partial_t \phi_r)^3 Y_0 = \partial_\mu V^{\mu}_{(0,2)}
		\, .
	\end{aligned}
\end{equation}
with $V_{(0,i)}$, where $i$ indicates the number of time derivatives. 
At order one in the $a$ field, the conditions are given by
\begin{equation}
	\begin{aligned}
		& E_1 +  2i\beta X_0 \partial_t \phi_r = -E_1 ,\\ &
		E_2 + 2 i \beta X_1 \partial_t \phi_r -3 \beta^2 (\partial_t \phi_r)^2 Y_0 = E_2 \, , 
	\end{aligned}
\end{equation}
while at the second order
\begin{equation}
	\begin{aligned}
		X_1 + 3 i \beta Y_0 \partial_t \phi_r = - X_1 \, .  
	\end{aligned}
\end{equation}
The independent conditions are
\begin{equation}
	\begin{aligned}
		& X_1  = -\frac{3i \beta}{2}  (\partial_t \phi_r)  Y_0 , \\ &  
		E_1 = i  (\partial_t \phi_r) \beta X_0 , \\
		& i \beta  E_0 \partial_t \phi_r  = \partial_\mu V^{\mu}_{(0,0)}, \\ & 
		\partial_\mu V^{\mu}_{(0,1)} = 0,  \\ &
		i\beta E_2  \partial_t \phi_r + \frac{i}{2} \beta^3 (\partial_t \phi_r)^3 Y_0 = \partial_\mu V^{\mu}_{(0,2)}
		\, .
	\end{aligned}
\end{equation}
In the previous section, we discussed the implications of 
\begin{equation}
	\begin{split}
		&E_1 = i  (\partial_t \phi_r) \beta X_0 , \\
		& i \beta  E_0 \partial_t \phi_r  = \partial_\mu V^{\mu}_{(0,0)}, \\ & 
	\end{split}
\end{equation}
that contribute to leading order where
and $E_0= \frac{\delta \mathcal{H}}{\delta \phi_r}$. 

The second order is specified by two additional independent coefficients $Y_0$ and $E_2$. The resulting action is 
\begin{equation}
	\Gamma[\Phi] = \int_{t,\boldsymbol{x} }  \Big \{ E_2 \phi_a + Y_0 \Big(  -\frac{3 i \beta}{2}\left(\partial_t \phi_r\right)\phi_a^2 +\phi_a^3  \Big) \Big \}\, ,
\end{equation}
The equation  
\begin{equation}
	i\beta E_2  \partial_t \phi_r + \frac{i}{2} \beta^3 (\partial_t \phi_r)^3 Y_0 = \partial_\mu V^{\mu}_{(0,2)}
	\, ,
\end{equation}
can be solved by noting that if one takes 
\begin{equation}
	V^{\mu}_{(0,2)} = (Z_t(\phi_r)(\partial_t\phi_r)^2 ,0) \, ,
\end{equation}
its divergence leads to 
\begin{equation}
	\partial_{\phi_r} Z_t (\partial_t\phi_r)^3 + 2 Z_t \partial_t \phi_r \partial^2_t \phi_r \, .
\end{equation}
Therefore, one can equate the coefficients to obtain
\begin{align}
	E_2 =  - i \frac{2}{\beta}  Z_t \partial^2_t \phi_r \, ,  \\ 
	Y_0  = -2 i \frac{1}{\beta^3} \partial_{\phi_r} Z_t \, ,
\end{align}
which leaves us with only one independent function of the field. 
The action in terms of $Z_t$ and its derivative is 
\begin{equation}
	\begin{aligned}
		\Gamma[\Phi] &= \int_{t,\boldsymbol{x} }  \Big \{- i \frac{2}{\beta}  Z_t \partial^2_t \phi_r \phi_a \\ &-2 i \frac{1}{\beta^3} \partial_{\phi_r} Z_t \Big(
		\phi_a^3  -\frac{3 i \beta}{2}\left(\partial_t \phi_r\right)\phi_a^2   
		- \frac{1}{2} \beta^2 (\partial_t \phi_r)^2\phi_a
		\Big)  \Big \}\, .
	\end{aligned}
\end{equation}
 If the spatial gradients are included, then additional terms will appear.



\section{Flow equations}
\label{app:flows}
We work in the Keldysh basis, see \labelcref{eq:Keldysh_basis} for our basic setup and, e.g.,~\cite{Berges:2012ty, Gasenzer:2007za, Gasenzer:2010rq, Corell:2019jxh, Berges:2008sr, Sieberer:2015svu} for a comprehensive introduction in the context of the fRG or \cite{Liu:2018kfw} for a more general perspective.
To keep the notation readable, we suppress the explicit RG scale $k$ dependence; it is implicitly understood everywhere.
In this convention, the flow of $\Gamma$ under the RG scale $k$ is given by the Wetterich equation~\cite{Wetterich:1992yh} which reads
\begin{equation}
\label{eq:wetterich}
\begin{aligned}
	\dot{\Gamma}[\Phi]&=   \frac{i}{2} \tr \Big[  \dot{R} \cdot (\Gamma^{(2)}[\Phi]+R)^{-1} \Big]  \\ &
	= \frac{i}{2} \int_{x,y} \dot R_{ij} (x-y)\Big[\Gamma^{(2)}[\Phi] +R\Big]^{-1}_{ji}(y,x)
	\, ,
\end{aligned}
\end{equation}
where the dot represents a derivative with respect to the logarithm of the RG scale,  
$R$ denotes the regulator and $\Gamma^{(2)}$ represents the two-point function. 
In general, the $n$-point vertex functions are obtained via the  functional derivatives of $\Gamma$, defined as
\begin{align}
\label{eq:gamma_moments}
\Gamma_{\alpha_{1}\ldots\alpha_{n}}^{(n)}(x_1,\ldots , x_n) = \frac{\delta^n \Gamma[\Phi]}{\delta\phi_{\bar{\alpha}_1}(x_1) \ldots \delta\phi_{\bar{\alpha}_n}(x_n)} 
\,,
\end{align}
where the $\alpha_i\in\{r, a\}$ indicate either a retarded or advanced index and a bar denotes index conjugation, i.e., $\bar{a} = r$ and $\bar{r} = a$.

For simplicity, we restrict ourselves to a frequency-independent regulator with only nonvanishing off-diagonal components as
\begin{align}
R = \begin{pmatrix}
	0 & R_{ar}(\boldsymbol{p}) \\
	R_{ra}(\boldsymbol{p}) & 0
\end{pmatrix}
\, .
\end{align}
To maintain the causal structure of the Keldysh action,  the retarded and advanced parts are connected by complex conjugation, i.e., $R_{ar}(\boldsymbol{p}) = R_{ra}^*(\boldsymbol{p})$.
As the regulator, we chose the standard spatial Litim regulator
\begin{equation}
R_{ar}(\boldsymbol{p}) = 	R_{ra}(\boldsymbol{p}) = r(\boldsymbol{p}) =  (k^2 -\boldsymbol{p}^2) \theta(k^2 -\boldsymbol{p}^2)
\, .
\end{equation}
This choice of regulator is compatible with the fluctuation-dissipation relation.

\subsection{Correlation functions and their flow}
Correlation functions can be obtained via functional derivatives of the generating functional.
For example, the matrix of connected two-point correlation functions is given by

\begin{equation}
\begin{aligned}
	i G\left(x, x^{\prime}\right) &  \equiv\left(\begin{array}{ll}
		\left\langle\phi_r(x) \phi_r\left(x^{\prime}\right)\right\rangle & \left\langle\phi_r(x) \phi_a\left(x^{\prime}\right)\right\rangle \\
		\left\langle\phi_a(x) \phi_r\left(x^{\prime}\right)\right\rangle & \left\langle\phi_a(x) \phi_a\left(x^{\prime}\right)\right\rangle
	\end{array}\right) \\ & =
	\left(\begin{array}{cc}
		i	G_{rr}\left(x, x^{\prime}\right) & iG_{ra}\left(x, x^{\prime}\right) \\
		i	G_{ar}\left(x, x^{\prime}\right) & 0
	\end{array}\right).
\end{aligned}
\end{equation}
The propagator is, in compact matrix notation, given by
\begin{equation}
\label{propagator}
G\left[\Phi\right]=-\left(R +\Gamma^{(2)}\left[\Phi \right]\right)^{-1} .
\end{equation}
We are left with the projection of the flow onto the scale-dependent couplings in the truncated quantum effective action \labelcref{eq:effGamma}.
We start by looking at the flow of the (off-shell) one-point function 
\begin{equation}
\begin{aligned}
	\label{eq:one_deriv_Gamma}
	&	 \dot \Gamma_b^{(1)}(z)  = \frac{\delta }{\delta \phi_b(z)}\partial_k \Gamma \\&
	=  -\frac{i}{2}  \int_{x,y,w,o} 
	\dot R_{ j k}(x,y)G_{k l}(y,w)
	\Gamma_{b l m}^{(3)}(z, w, o) G_{m j}(o, x)
	\, .
\end{aligned}
\end{equation}
with $b = a,r$, where the three-point function is defined as 
\begin{equation}
{\Gamma}_{b l m}^{(3)}(r, x,y )=\frac{\delta {\Gamma}^{(2)}_{l m}(x,y)}{\delta \phi_b(r)}.
\end{equation}
Hereby, one requires the functional derivative of the propagator with respect to the field
\begin{equation}
\begin{aligned}
	&\frac{\delta}{\delta \phi_{i}\left(x \right)} G_{bc}(y, z) =
	\\ &
	-  \int_{x',y'} G_{b l }\left(y,y'\right) \Gamma_{i l m}^{(3)}\left(x,y',x' \right) G_{ m c}\left(x', z\right)
	\, .
\end{aligned}
\end{equation}
Assuming that the physical system under study is translationally invariant in space, it is straightforward to derive the flow equation of the effective potential by performing the integral over the internal frequency after the matrix multiplications
\begin{equation}
\label{eq:flow_effpot}
\begin{aligned}
	\partial_k U^{(1)} = & - \frac{1}{2}\int_{\boldsymbol{p}} \dot{r}(\boldsymbol{p}^2) \frac{U^{(3)}}{(U^{(2)}+\boldsymbol{p}^2 + r(\boldsymbol{p}^2) )^2 }
	\, ,
\end{aligned}
\end{equation}
after frequency integration.

Putting all the pieces together, we arrive at the flow for the derivative of the effective potential, given in~\labelcref{eqforV}.

The flow equation for the equilibration rate $X$ is obtained by projecting onto the flow of the two-point function and similarly for the four-point function.
The flow equation explicitly reads
\begin{equation}
\begin{aligned}
	& \dot  \Gamma_{a b}^{(2)}(q, -q)= i  \int_p \Big\{ \dot R_{i j}(p)\Big(G_{j k}(p) \\
	& \times \Gamma_{ a k l}^{(3)}(q, p,-q-p) G_{l m}(p+q) 
	\Gamma_{ b m n}^{(3)}\left(-q, p+q,-p\right) G_{ n i}\left(p\right) \\
	& - \frac{1}{2} G_{ j k}(p) \Gamma_{a b k l}^{(4)}(q,-q, p,-p) G_{ l i}(p)\Big) \Big\}
	\, ,
\end{aligned}
\end{equation}
where $q=(\nu, \boldsymbol{q})$ is the external momentum.
Now we want to specify our ansatz for the effective action.
The only nonvanishing components of the three-point function are given by
\begin{equation}
\begin{aligned}
	&  \Gamma_{ara}^{(3)}(p,q, r)=  \Gamma_{aar}^{(3)}(p,r, q) \\ & = (2 \pi)^{-2(d+1)}  \delta^{d+1}\left(p+q+r\right) 2 i X^{(1)},
\end{aligned}
\end{equation}
and
\begin{equation}
\Gamma_{rrr}^{(3)}(p,q, r)=(2 \pi)^{-2(d+1)}  \delta^{d+1}\left(p+q+r\right) \gamma_3(p,q),
\end{equation}
where
\begin{equation}
\gamma_3\left(p, q\right)= U^{(3)} -i X^{(1)}\left( \omega_p+ \omega_q\right). 
\end{equation}
Similarly, the only nonvanishing components of the four-point function are given by
\begin{equation}
\Gamma_{rrra}^{(4)}(p,q, r,s)=(2 \pi)^{-2(d+1)}  \delta^{d+1}\left(p+q+r+s\right) \gamma_4(p,q,r),
\end{equation}
and
\begin{equation}
\Gamma_{rraa}^{(4)}(p,q, r,s)=(2 \pi)^{-2(d+1)}  \delta^{d+1}\left(p+q+r+s\right) 2 i X^{(2)},
\end{equation}
where
\begin{equation}
\begin{aligned}
	\gamma_4\left(p, q, r\right)= U^{(4)} -iX^{(2)} \left( \omega_p+ \omega_q+ \omega_r\right).
\end{aligned}
\end{equation}
The frequency integration can be performed analytically, and we are only left with an integral over the spatial loop momentum $p$

\begin{align}	
\label{eq:flow_}
&\partial_k X  =  \frac{1}{\text{Vol}_{d+1}}\lim_{\nu\to0} \partial_{i\nu} \left[ \lim_{\boldsymbol{q}\to0} \partial_{t}\Gamma_{ra}^{(2)}(q,-q) \right] = \nonumber \\
&  \frac{1}{2}\int_{\boldsymbol{p}} \frac{\partial_t r(\boldsymbol{p}^2)}{(U^{(2)} + 
	\boldsymbol{p}^2 + r(\boldsymbol{p}^2))^2}\Bigg[3\frac{(U^{(3)})^2}{(U^{(2)} + 
	\boldsymbol{p}^2 + r(\boldsymbol{p}^2))^2} X \nonumber \\
&-\frac{  8 U^{(3)} X^{(1)}}{ (U^{(2)} + 
	\boldsymbol{p}^2 + r(\boldsymbol{p}^2))}+ 2 X^{(2)}\Bigg]
\, .
\end{align}
 Being related to a transport coefficient, the plasmon limit is the correct choice for the equilibration rate $X$. The final equation is given in~\labelcref{eqforX}.

\subsection{ Flow equations at finite spatial volume}
\label{Finite_scaling}

In this subsection, we briefly overview how to modify the fRG flows to suit a finite spatial volume. For a more comprehensive understanding, we recommend referring to,
e.g.,~\cite{Braun:2005gy, Braun:2011uq, Braun:2011iz, Tripolt:2013zfa, Fister:2015eca}.
As commonly employed, we use a box with an extension of $L$ in all spatial directions, denoted by $i=1, ...,d$, where $x_i\in \left[0, L\right]$, and apply periodic boundary conditions.
A finite extent in a given direction with periodic boundary
conditions $\phi(x+L) = \phi(x)$ only allows for plane waves that are periodic under shifts $x_i \rightarrow x_i+L$, that are $\exp \left(i 2 \pi n x_i / L\right)$ with $n \in \mathbb{Z}$. 
This leads to a discrete set of  momentum modes given by
\begin{equation}
\label{eq:fin_vol_mom_modes}
p_i= \frac{2 \pi n_i}{L}, \quad \text { with } \quad n_i \in \mathbb{Z} 
\, ,
\end{equation}
for all spatial directions.
Consequently, each spatial momentum
integral gets replaced by a sum
\begin{equation}
\int_{-\infty}^{\infty} d p_i \rightarrow \frac{2 \pi}{L} \sum_{n_i \in \mathbb{Z}} \quad .
\end{equation}
The flow equations for a system
in a finite spatial volume $L^d$ can be implemented efficiently by introducing the mode-counting function $\mathcal{B}_d$ as
\begin{equation}
\label{eq:conting_function}
\mathcal{B}_d(kL)=\frac{(2\pi)^d}{L^d}\sum_{n}   \Theta \left(\left(\frac{ kL}{2\pi}\right)^2-	L^2 \vec{n}^2  \right)
\, ,
\end{equation}
where we use the shorthand notation
\begin{equation}
\vec{n}^2=\frac{4 \pi^2}{L^2} \sum_{i=1}^d n_i^2
\, .
\end{equation}
The Heaviside step function $\Theta$ ensures that only modes with a magnitude less than $k$ contribute to the sum.
In the limit of infinite-volume, $(L \to \infty)$ the infinite-volume result is recovered

\begin{equation}
\lim _{k L \rightarrow \infty} \mathcal{B}_d(k L)=	\Omega_d \frac{k^d}{d}
\, .
\end{equation}
For the derivative of the effective potential, the flow equation at a finite volume with length $L$ results in
\begin{equation}
\label{eq:flow_effpot_fin_vol}
\partial_k U^{(1)} = \frac{\partial}{\partial \phi} \left( \frac{\mathcal{B}_d(kL)}{(2 \pi)^d} k  \frac{1}{U^{(2)}+ k^2  } \right)
\, .
\end{equation}
For the dissipation rate, the result is
\begin{equation}
\label{eq:flow_X_fin_vol}
\partial_k X =- \frac{\mathcal{B}_d(kL) k}{2 (2\pi)^d}\Bigg[3 (\partial_\phi G)^2 X + 4\partial_\phi (G^2)X^{(1)}+ 2 G^2X^{(2)}\Bigg]
\, .
\end{equation}

\section{Numerical algorithm}
\label{sec:num_alg}
The explicit expressions for the flow of the derivative of the effective potential $U^{(1)}(\phi)$ and the dissipation rate $X(\phi)$ are given by nonlinear partial differential equations in the field variable; see \labelcref{eqforV} and \labelcref{eqforX}, respectively.
Solving these numerical equations requires some attention to detail due to the nonanalyticity of the solutions. Here, we follow the general ideas introduced in \cite{Grossi:2019urj}, which have been developed further in \cite{Koenigstein:2021syz, Ihssen:2022xkr, Ihssen:2023qaq}.
We briefly recapitulate how the upwinding method can stabilize spatial discretization for the reader's convenience.

In the case of infinite-volume, we use $N=800$ stencils in field space with a uniform spacing (introduced below) of $d\phi=0.02$. For finite volume, we use $N=400$. We have checked the stability against varying these parameters.

Without loss of generality, we can restrict the discussion to the case $\phi > 0$.
In this case, the notion of upwind derivatives is fixed entirely by the direction of the derivative.

The boundary condition at large field values is fixed via extrapolation with a ghost cell.
We use (anti)symmetry around zero at the vanishing field value to fix the boundary conditions.

\subsection{Derivative of the potential}

The flow equation for the derivative of the potential here, abbreviated as $u(t, \phi) = U^{(1)}(t,\phi)$, where now $t$ is the RG time, can be expressed formally as
\begin{equation}
\label{eqforu}
\partial_t u = A_d \partial_\phi f
\, , 
\end{equation}
where $A_d = \frac{\Omega_d}{(2 \pi)^d}\frac{k^{d+2}}{d}$ and $f(t,u)$ [see \labelcref{definitionG}] represents the flux of a single scalar field.
Here, we aim at a linear upwind scheme to discretize \labelcref{eqforu}, already given in a conservative form.
Following the linear upwind scheme in \cite{Ihssen:2023qaq}, we denote cells elements by  $[\phi_{i-\frac12}, \phi_{i+\frac12} ]$, with the size of the $i$th cell $\mathrm{d}\phi_i =  \phi_{i+\frac12}-  \phi_{i-\frac12}$.
Later on, we will also need the second derivative of the potential, given by $q(t, \phi) = \partial_\phi u(t, \phi)$.

The first equation \labelcref{eqforu} takes the form of a transport equation with a constant negative speed, thus requiring a right derivative to incorporate the direction of transport
\begin{equation}
\partial_t u_i+  A_d\frac{f_{i+1}-f_{i}}{\mathrm{d} \phi_i} = 0
\, .
\end{equation}
The dissipative behavior is recovered by discretizing the auxiliary equation for $q(t, \phi)$ with a left derivative
\begin{equation}
q_{i}^L= \frac{u_{i} - u_{i-1}}{\mathrm{d} \phi_i}
\, .
\end{equation}
The resulting discretization is 
\begin{equation}
\partial_t u_i =  A_d\frac{f^{R}_i - f^{L}_i}{\mathrm{d}\phi_i}
\, ,
\end{equation}
where the fluxes are chosen according to an alternating pattern as
\begin{equation}
\begin{aligned}
	& f_i^R= f\left(t, \frac{u_{i+1}- u_{i}}{\mathrm{d}\phi_i } \right) \, , \\
	& f_i^L= f\left(t, \frac{u_{i}- u_{i-1}}{\mathrm{d}\phi_i } \right)
	\, .
\end{aligned}
\end{equation}

\subsection{Relaxation rate}

Due to the exponential growth of the dissipation rate $X$, we switch variables to the logarithm thereof $\tau~=~\log{X}$.
Its flow, cf.~\labelcref{eqforX}, is given by
\begin{equation}
\begin{aligned}
	\label{eqfortau}
	\partial_t \tau &= \alpha(\partial_\phi f)^2  + \beta (\partial_\phi f^2) (\partial_\phi \tau) \\
	&+ \gamma f^2(\partial^2_\phi \tau + (\partial_\phi \tau)^2 )
	\, ,
\end{aligned}
\end{equation}
where $\alpha$, $\beta$ and $\gamma$ are time-dependent coefficients. 
This equation is a nonlinear advection-diffusion equation with a negative advection coefficient $\sim$$\beta$ and a positive diffusion $\sim$$\gamma$. 
The quantity $f$ is computed using a discretization that is consistent with the derivative of the potential, i.e., using a right derivative, such that the first term is given by
\begin{equation}
\alpha(\partial_\phi f)^2  \to \alpha\Big(  \frac{f_{i}^R- f_i^L}{\mathrm{d}\phi_i}\Big)^2
\, .
\end{equation}
The advection term is replaced by a first-order left approximation of the upwind first derivative according to \cite{osher1982upwind}
\begin{equation}
\beta \partial_\phi f^2 \partial_\phi \tau\to\beta \frac{(f_{i}^R)^2- (f_i^L)^2}{\mathrm{d}\phi_i}\frac{ \tau_{i}-\tau_{i-1}}{\mathrm{d}\phi_i }
\, .
\end{equation} 

Since  we are dealing with an upwinding discretization of an advection-diffusion equation, the second derivative of the diffusion term is discretized using a standard central derivative
\begin{align}
\gamma f^2(\partial_\phi \tau)^2  &\to  \gamma f_i^2\Bigg(  \left( \frac{\tau_i-\tau_{i-1}}{\mathrm{d}\phi_i} \right)^2 \Bigg) , \\
\partial^2_\phi \tau   &\to  \frac{\tau_{i+1}-2\tau_i +\tau_{i-1}}{\mathrm{d}\phi_i^2 }
\, .
\end{align}

This concludes our discussion of the spatial discretization. The modifications in the equations when considering the system in a finite volume, cf. Appendix~\ref{Finite_scaling}, do not alter the structure of the equations. Hence, the discretization is the same.

\clearpage
\newpage
\bibliography{bib_master}

\end{document}